\documentclass[aps,prl,showpacs,notitlepage,singlecolumn,superscriptaddress,11pt,floatfix,
]{revtex4-1}
\usepackage{amsmath}
\usepackage{amssymb}
\usepackage{amsfonts}
\usepackage{stmaryrd}
\usepackage{wasysym}
\usepackage{graphicx}
\usepackage{multirow}
\usepackage{tabularx}
\usepackage{textcomp}
\usepackage{subfigure}
\usepackage{url}
\usepackage[colorlinks=true,citecolor=blue,urlcolor=blue]{hyperref}
\usepackage{romannum}
\usepackage{mathtools}
\usepackage[utf8]{inputenc}
\usepackage{braket}
\usepackage{amsmath}
\usepackage{xcolor}
\usepackage{colortbl}
\usepackage{color,soul} 
\usepackage{bm}
\usepackage[normalem]{ulem}
\usepackage[utf8]{inputenc}
\usepackage{array}
\usepackage{makecell}
\usepackage{multirow}
\usepackage{enumitem}  
\DeclareRobustCommand{\rchi}{{\mathpalette\irchi\relax}}
\newcommand{\irchi}[2]{\raisebox{\depth}{$#1\chi$}} 
\usepackage{lineno}

\hypersetup{colorlinks=true, urlcolor=blue, citecolor=cyan, pdfborder={0 0 0}}

\pagecolor{white}

\usepackage{fancyhdr}
\usepackage{pdfpages} 
\usepackage{pgffor} 

\makeatletter
\AtBeginDocument{\let\LS@rot\@undefined}
\makeatother
\def\supplementfilename{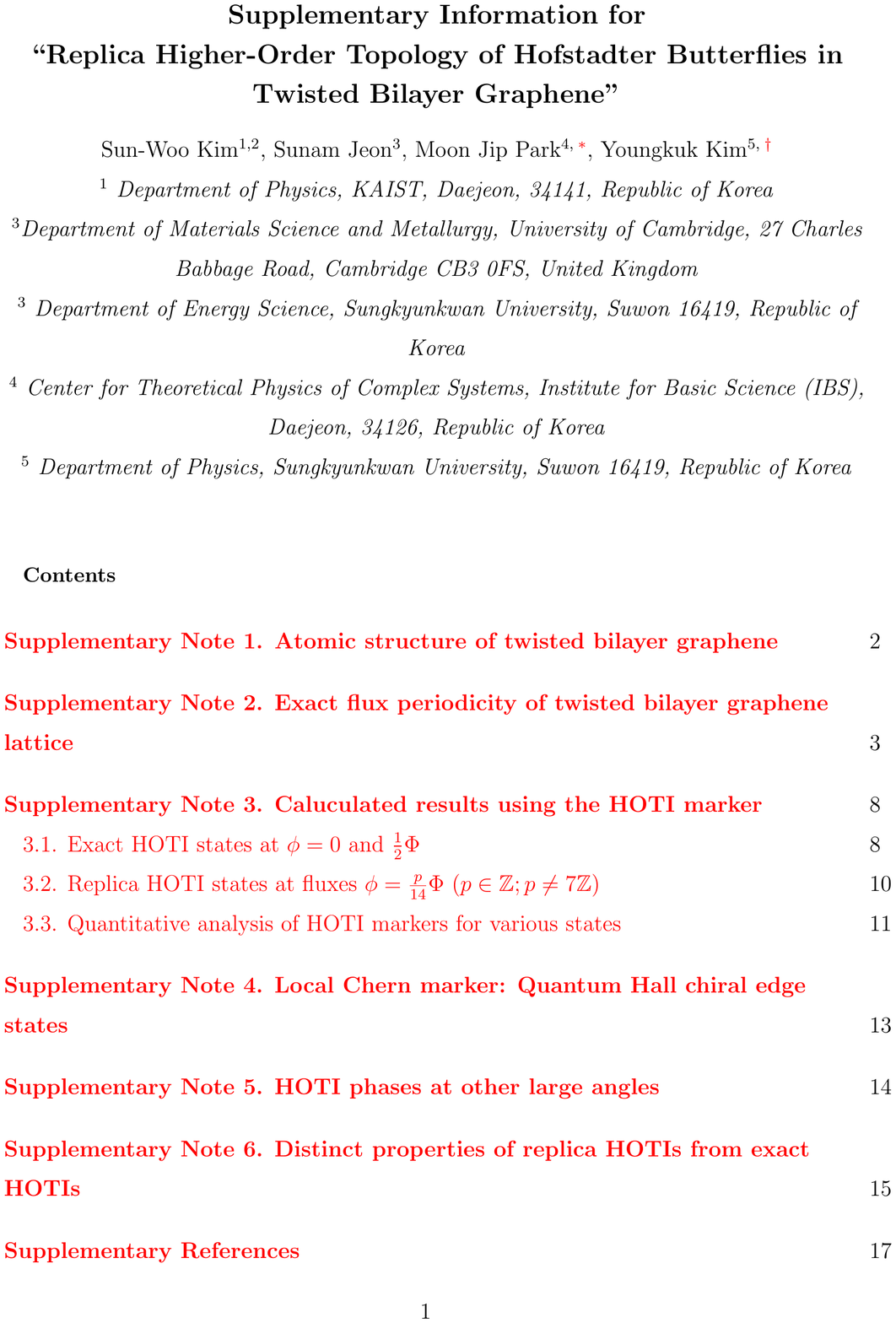}

\pdfximage{\supplementfilename}
\def\numbersupplementpages{\the\pdflastximagepages}

\newif\ifarXiv
\arXivtrue 

\begin{document}
\pagenumbering{arabic}

\title{Replica Higher-Order Topology of Hofstadter Butterflies in Twisted Bilayer Graphene}

\author{Sun-Woo Kim,$^{1,2}$ 
Sunam Jeon,$^{3}$ 
Moon Jip Park,$^{4,5,\textcolor{red}{*}}$ and 
Youngkuk Kim$^{6,\textcolor{red}{\dagger}}$ \\
$^{1}$ {\it Department of Physics, KAIST, Daejeon, 34141, Republic of Korea}\\
$^{2}$ {\it Department of Materials Science and Metallurgy, University of Cambridge, 27 Charles Babbage Road,
Cambridge CB3 0FS, United Kingdom}\\
$^{3}$ {\it Department of Energy Science, Sungkyunkwan University, Suwon 16419, Republic of Korea}\\
$^{4}$ {\it Center for Theoretical Physics of Complex Systems, Institute for Basic Science (IBS), Daejeon, 34126, Republic of Korea}\\
$^{5}$ {\it Department of Physics, Hanyang University, Seoul 04763, Republic of Korea}\\
$^{6}$ {\it Department of Physics, Sungkyunkwan University, Suwon 16419, Republic of Korea}\\
$^{\textcolor{red}{*}}$
 {\it \href{mailto:moonjippark@hanyang.ac.kr}{\textcolor{blue}{$\mathrm{moonjippark@hanyang.ac.kr}$}}}
\\
$^{\textcolor{red}{\dagger}}$
 {\it \href{mailto:youngkuk@skku.edu}{\textcolor{blue}{$\mathrm{youngkuk@skku.edu}$}}}
} 

\date{\today}

\maketitle

\section{Abstract}
The Hofstadter energy spectrum of twisted bilayer graphene (TBG) is found to have recursive higher-order topological properties. We demonstrate that higher-order topological insulator (HOTI) phases, characterized by localized corner states, occur as replicas of the original HOTIs to fulfill the self-similarity of the Hofstadter spectrum.  We show the existence of exact flux translational symmetry in TBG at all commensurate angles. Based on this result, we identify that the original HOTI phase at zero flux is re-entrant at a half-flux periodicity, where the effective twofold rotation is preserved. In addition, numerous replicas of the original HOTIs are found for fluxes without protecting symmetries. Like the original HOTIs, replica HOTIs feature both localized corner states and edge-localized real-space topological markers. The replica HOTIs originate from the different interaction scales, namely, intralayer and interlayer couplings, in TBG. The topological aspect of Hofstadter butterflies revealed in our results highlights symmetry-protected topology in quantum fractals. 

\clearpage
\section{Introduction}
Magnetic translational symmetry of crystals in the presence of an external magnetic field \cite{Zak64pA1602, Zak64pA1607} manifests as a fractal form of the energy spectrum that resembles recurring replicas of butterflies, known as Hofstadter butterflies \cite{PhysRevB.14.2239,harper1955single, azbel1964energy, langbein1969tight, Claro79p6068, PhysRevLett.86.147, PhysRevLett.92.036802}.  Although a strong magnetic field is generally required, Hofstadter butterflies have recently been observed owing to advances in two-dimensional van der Waals materials \cite{bistritzer11p12233, bistritzer11p035440, geim2013van, Chen14p075401, Ferrari2015,Novoselov2016, cao18p80, cao18p43, Balents2020}. The magnetic field required to produce replicas of the Landau levels could be significantly reduced by the large-scale synthesis of a van der Waals superlattice with a macroscopic unit cell. For this crucial development, the Hofstadter butterflies have been experimentally realized in a graphene superlattice \cite{Dean2013, Ponomarenko2013, Hunt2013, Wang2015, Yang2016,Spanton2018}, magic-angle twisted bilayer graphene (TBG) \cite{Lu2021, Saito2021}, and twisted double-bilayer graphene \cite{arxiv.2006.14000}.

Notably, the link with the magnetic translational symmetry and symmetry-protected topological phases of matter has been revealed recently \cite{Otaki19p245108, Wang20p236805, herzog2020hofstadter, Lian20p041402, guan2021landau, das2022observation, herzog2021reentrant, herzog2022magnetic, bartholomew2020fractional, Zuo_2021}. 
In a general lattice model with multiple sites per unit cell, the Hofstadter energy spectrum becomes approximately replicative under the addition of the flux periodicity, $E(\phi)\approx E(\phi+\Phi)$, which constitutes the additional flux translational symmetry via the unitary transformation of the Hamiltonian, $H(\phi)$, as,
\begin{eqnarray}
\mathcal{U}(\mathbf A) H(\phi)\,\mathcal{U}^\dagger(\mathbf A) \approx H(\phi  + \Phi), 
\end{eqnarray}
where $\mathcal{U}(\mathbf A)=\sum_{\mathbf{R}} c^\dagger_{\mathbf{R}}c_{\mathbf{R}}\exp{(i\frac{e}{\hbar}\int_{\mathbf{r}_0}^{\mathbf{R}}{\mathbf {A}}\cdot d\mathbf{r})}$ with $\int(\nabla\times{\mathbf {A}}) \cdot d\mathbf{S}=\Phi$, $c^\dagger_\mathbf R$ ($c_\mathbf R$) is a creation (annihilation) operator of an electron at $\mathbf R$ in the real space, $\mathbf A$ is the vector potential, and $S$ is the unit cell area.
Remarkably, the effective time-reversal symmetry $\mathcal{UT}$ is restored at a half-flux periodicity $\phi= \frac{1}{2}\Phi$ \cite{herzog2020hofstadter}, allowing for the existence of diverse topological states of matter protected by symmetries  \cite{Otaki19p245108, Wang20p236805, herzog2020hofstadter, Lian20p041402, guan2021landau}.

\begin{figure}[h]
\includegraphics[width= \textwidth]{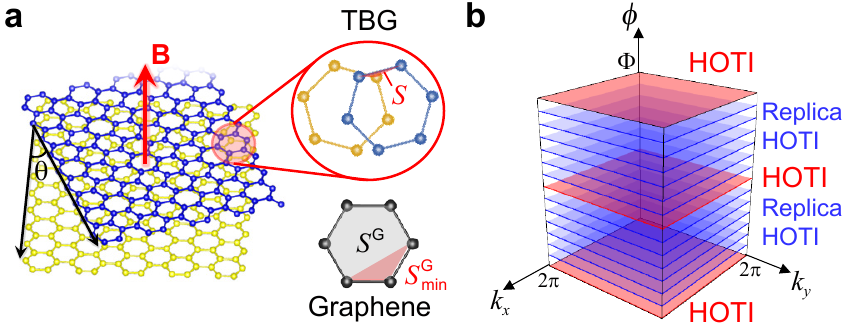} 
\caption{
\label{fig:replica}
Replica HOTI states in TBG.
\textbf{a} Atomic structure of TBG. 
$S$ denotes the smallest area enclosed by a Peierls path for TBG at $\theta$=21.8$^{\circ}$.
For comparison, we illustrate the corresponding area for graphene $S^\textrm{G}_{\textrm{min}}$ and the area of the graphene unit cell $S^\textrm{G}$, where $S =\frac{1}{7} S^\textrm{G}_{\textrm{min}} = \frac{1}{42} S^\textrm{G}$.
\textbf{b} Schematic of re-entrant exact HOTI and replica HOTI phases as a flux $\phi$ function.
Red and blue colors indicate the exact HOTI and replica HOTI phases, respectively.
The $k_x$ and $k_y$ are pseudo-momenta, well defined in the presence of the magnetic translational symmetry under arbitrary rational flux.
}
\end{figure}

In this work, we study the higher-order topological insulator (HOTI) phases of Hofstadter butterflies in TBG. Archetypal HOTIs have been studied with respect to symmetry protection \cite{Otaki19p245108, herzog2020hofstadter}.  By contrast, the replica HOTIs that we find here recur in the form of quasiperiodic replicas without explicit symmetry protection. Instead, they rely on the self-similar nature of Hofstadter butterflies. We prove that the full lattice model of TBG possesses the exact flux periodicity at all commensurate angles, which rigorously characterizes the band topological protection in the presence of the magnetic field. Two original HOTIs exist at time-reversal invariant fluxes (TRIFs) $\phi = 0$ and $\phi = \frac{1}{2}\Phi$, where $\phi = -\phi$ (mod  $\Phi$). In addition, replicas HOTIs recur at the specific fluxes $\phi=\frac{p}{2N_{\textrm{rep}}}\Phi ~(p\in\mathbb{Z}; p\neq N_{\textrm{rep}}\mathbb{Z}$) [Fig.\,\ref{fig:replica}\textbf{b}; See Eq. \eqref{eq:nrep} for the definition of $N_{\textrm{rep}}$]. To quantitatively diagnose HOTIs, we extend the concept of real-space topological markers \cite{bianco2011mapping, Shem2014topological, tran2015topological, caio2019topological, mondragon2019robust,varjas_2020_computation} to the HOTI version.  Similar to the original HOTIs, replica HOTIs are characterized by localized HOTI markers and corner modes. The origin of the replica HOTIs is attributed to the reduced interior area of the Peierls path because of the interlayer hopping in TBG. 

\section{Results and discussion}

\subsection{Lattice model and symmetries}
We use the Moon-Koshino tight-binding model for TBG \cite{Moon12p195458}
\begin{eqnarray}
H=\sum_{ij} t_{ij}(\mathbf{R}_i-\mathbf{R}_j) c^\dagger_{\mathbf{R}_i}c_{\mathbf{R}_j} + h.c.,
\end{eqnarray}
where the hopping integral $t_{ij}(\mathbf{R}_i-\mathbf{R}_j)$ is modelled as an exponentially decaying function of $\mathbf{R}_i-\mathbf{R}_j$~\cite{Moon12p195458} (see Methods). Magnetic flux $\phi$ is introduced using the Peierls substitution\,\cite{peierls1933theorie}: $t_{ij}(\mathbf{R}_i-\mathbf{R}_j) \rightarrow t_{ij}(\mathbf{R}_i-\mathbf{R}_j)\exp{\left(i\frac{e}{\hbar}\int_{\mathbf{R}_j}^{\mathbf{R}_i} \mathbf{A}_0\cdot d\mathbf{r}\right)}$, 
where $\mathbf{A}_0$ is the vector potential in the Landau gauge (see Methods).
We consider the atomic structure of TBG in the hexagonal space group \#\,177, generated by twisting the AA-stacked bilayer graphene about the hexagonal center with the twist angle $\theta_{m,n}=\arccos{\frac{1}{2}\frac{m^2+n^2+4mn}{m^2+n^2+mn}}$ ($m,n\in \mathbb{Z}, m\neq n$) (Fig.\,\ref{fig:replica}a). This construction of TBG preserves $C_{6z}$, $C_{2x}$, and $C_{2y}$ rotational symmetries. The twist lowers the discrete translational symmetry, leading to the translational symmetry of the moir\'{e} lattice with the enlarged unit cell area by $N_L=m^2+n^2+mn$ times. In the presence of a uniform perpendicular magnetic field, a flux translational symmetry emerges, which locally restores crystalline symmetries for specific fluxes.  For example, for $\phi=\frac{1}{2}\Phi$, combination of $C_{2x}$ and unitary matrix $\mathcal{U}$ leaves the system invariant.  Therefore, $\mathcal{U}C_{2x}$ is preserved at $\phi = \frac{1}{2}\Phi$ because $\frac{1}{2}\Phi = -\frac{1}{2}\Phi$ (mod $\Phi$). 

\subsection{Hofstadter butterflies}
For the nearest-neighbor tight-binding model of graphene, the flux periodicity is given as the magnetic field strength $B=\Phi_0/S^\textrm{G}$, where $\Phi_0=\frac{h}{e}$ is the flux quantum and $S^\textrm{G}$ is the graphene unit cell area \cite{bistritzer11p035440, Rhim12p235411, Moon12p195458}. However, when next neighbor hoppings are introduced, the minimal loop along the allowed hoppings, namely, the minimal Peierls path, has decreased inner area $S_{\textrm{min}}^{\textrm{G}}=\frac{1}{6}S^\textrm{G}$ (Fig.\,\ref{fig:replica}\textbf{a}), leading to an increased flux periodicity. A stronger magnetic field of $B=\Phi_0/{S^\textrm{G}_{\textrm{min}}}=6\Phi_0/{S^\textrm{G}}$ is required to implement the full flux quantum into the decreased inner area $S_{\textrm{min}}^{\textrm{G}}$ of the minimal Peierls path. Consequently, the entire cycle is completed by repeating six times modulated quasiperiodic replicas of the nearest-neighbor graphene spectrum \cite{herzog2020hofstadter}.

\begin{figure}[hb]
\includegraphics[width=\textwidth]{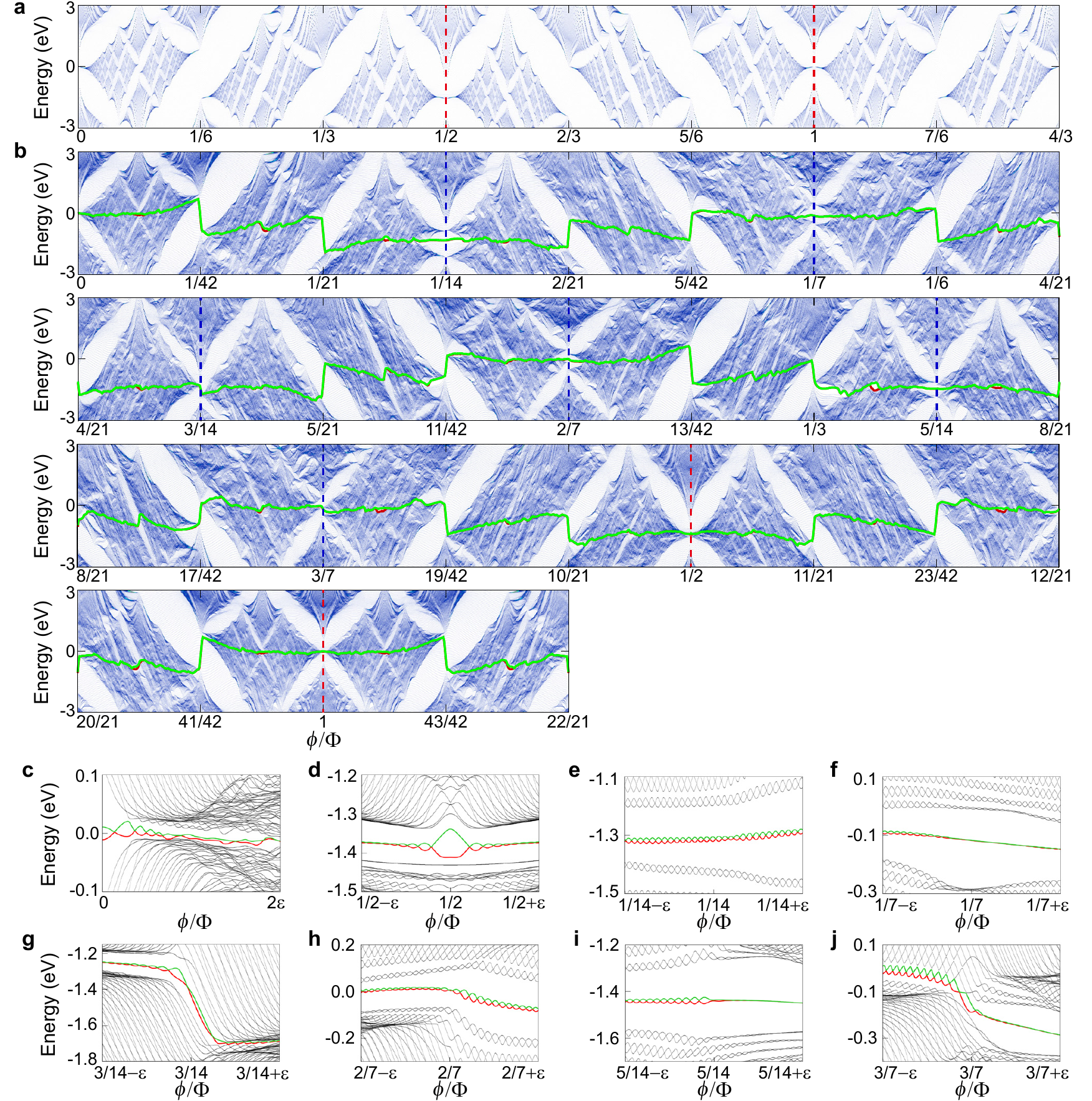} 
\caption{
\label{fig:energy}
Hofstadter butterflies of the atomistic tight-binding model of TBG.
\textbf{a-b} Hofstadter spectrum of TBG \textbf{a} without and \textbf{b} with interlayer coupling calculated by using the kernel polynomial method (see Methods). Red dashed lines indicate the system's half and full flux periodicity. Blue dashed lines indicate quasi periodicities at $\phi=\frac{p}{14}\Phi~ (p=1,2,\dots,6$). In \textbf{b}, the red and green solid lines denote the highest occupied and lowest unoccupied states, respectively. \textbf{c-j} Magnified view of \textbf{b} at  specific flux values \textbf{c} $\phi=0$, \textbf{d} $\phi=\frac{1}{2}\Phi$, \textbf{e} $\phi=\frac{1}{14}\Phi$, \textbf{f} $\phi=\frac{1}{7}\Phi$, \textbf{g} $\phi=\frac{3}{14}\Phi$, \textbf{h} $\phi=\frac{2}{7}\Phi$, \textbf{i} $\phi=\frac{5}{14}\Phi$, and \textbf{j} $\phi=\frac{3}{7}\Phi$, where $\varepsilon = \frac{1}{1680}\Phi$.}
\end{figure}

For TBG, we show the existence of the exact flux periodicity at the twist angle $\theta_{m,n}$, dictated by,
\begin{equation}
N_{\textrm{rep}}\equiv {S^\textrm{G}_{\textrm{min}}}/S_{\textrm{min}}^{\textrm{TBG}}=\frac{ N_L}{\textrm{gcd}(z_1,z_2,z_3)}\in \mathbb{Z},
\label{eq:nrep}
\end{equation}
where gcd indicates the greatest common divisor and $z_1=m^2-n^2$, $z_2=3m^2$, $z_3=2m^2-mn-n^2$ (see Supplementary Note 2).
For $\theta=21.8^{\circ}$($m=1$,$n=2$), $N_{\textrm{rep}}=7$ corresponds to the area of the minimal Peierls path, $S^{\mathrm{TBG}}_{\textrm{min}} \equiv S =\frac{1}{7} S^\textrm{G}_{\textrm{min}} = \frac{1}{42} S^\textrm{G}$ (see Fig.\,\ref{fig:replica}\textbf{a}).  As a result, a self-similar pattern is rendered by 42 replicas of the original graphene spectrum, only having the nearest-neighbor hopping term. 

Figure\,\ref{fig:energy} shows the calculated Hofstadter butterflies for both graphene and TBG by using the kernel polynomial method (see Methods).  Quasi-periodicity is exhibited, as our tight-binding model includes electron hopping beyond the nearest neighbors.   For example, in the graphene spectrum (Fig.\,\ref{fig:energy}\textbf{a}), the quasi-periodicity of $\frac{1}{6}\Phi$ is displayed by having similar patterns recurring at every integer multiple of $\frac{1}{6}\Phi$.  Similarly, for TBG spectrum (Fig.\,\ref{fig:energy}\textbf{b}), a quasi-periodicity of $\frac{1}{42}\Phi$ occurs as expected. Moreover, the energy spectrum that resembles the graphene spectrum in Fig.\,\ref{fig:energy}\textbf{a} recurs at every integer multiple of $\frac{1}{7}\Phi$.  This modulation of the graphene spectrum by $\frac{1}{7}\Phi$ is weaker than that of $\frac{1}{42}\Phi$ because the interlayer hopping is relatively weaker in TBG compared to next-nearest-neighbor intralayer hopping.  Therefore, the quasi-periodicity of $\frac{1}{7}\Phi$ is more prominent than that of $\frac{1}{42}\Phi$ in TBG spectrum.  

The computed spectrum exhibits symmetries of Hofstadter butterflies (Fig. \ref{fig:energy}\textbf{b}). Translational flux symmetry is displayed in the recurring patterns at $\phi$ and $\phi=\phi+\Phi$.  Moreover, the $C_{2x}$ symmetry that is broken under the flux gives rise to the mirror-symmetric spectrum about TRIFs (both $\phi=0$ and $\phi=\frac{1}{2}\Phi$). The Hamiltonian is transformed under the $C_{2x}$ operator as
\begin{eqnarray}
C_{2x}H(\phi)\,C_{2x}^\dagger = H(-\phi).
\end{eqnarray}
Combined with the unitary matrix $\mathcal{U}$, we obtain
\begin{eqnarray}
\mathcal{U}C_{2x} H\left(\tfrac{1}{2}\Phi + \phi\right)(\mathcal{U}C_{2x})^\dagger = H\left(\tfrac{1}{2}\Phi - \phi\right).
\end{eqnarray} 
Therefore, the energy eigenvalues for $\phi$ and $-\phi$ about TRIFs are equivalent. 

\subsection{Exact HOTIs}
\begin{figure}[h]
\includegraphics[width= \textwidth]{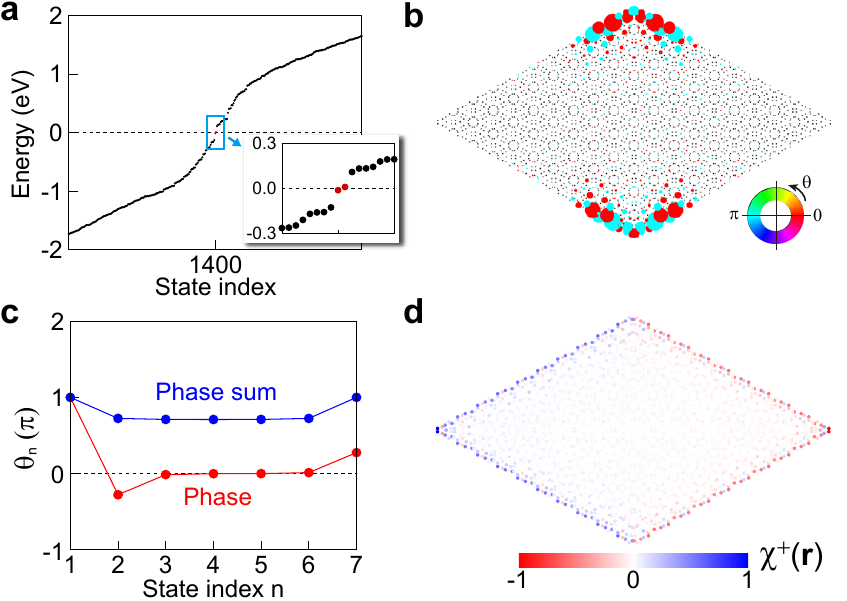} 
\caption{
\label{fig:hoti}
Characterization of HOTI states.
\textbf{a} Energy spectrum of TBG with $\theta=21.8^{\circ}$ in OBC.
Red Data points indicate corner states.
\textbf{b} Topological corner state in OBC.
The colored circle indicates the phases of eigenstate components.
\textbf{c} Zak phase calculated using $c_{2x}=+1$ bands along the rotation-invariant $k_y=0$ line in periodic boundary condition.
\textbf{d} HOTI marker $\rchi^{+}(\mathbf{r})$ for $c_{2x}=+1$ states in OBC.
Here, $\rchi^{+}(\mathbf{r})$ is normalized to its maximum value.
}
\end{figure}

The proposed tight-binding model reproduces the HOTI phase of TBG well at zero flux, showing good agreement with previous studies \cite{MJPark1, MJPark2}.  Consequently, the system harbors localized states at the corner of a diamond-shaped flake under an open boundary condition (OBC) (Fig.\,\ref{fig:hoti}\textbf{b}).  In energy space, two corner states reside inside the spectral gap of the bulk (Fig.\,\ref{fig:hoti}\textbf{a}).  In general, these two (in-gap) corner states can have different energies owing to the finite-size effect, in which they spatially overlap and cause hybridization \cite{MJPark2}. 

The bulk gap at $\phi=0$ is $\sim$ 9 meV. This leads to a spectral gap of approximately 236 meV for a flake width of $\sim 11$ nm (2800 atoms). The HOTI at $\phi=0$ is under multiple protection conditions \cite{MJPark1,MJPark2}. Two distinct topological invariants exist: the second Stiefel-Whitney number $\omega_2$ \cite{JAhn_2018_PRL,JAhn_2019,JAhn_2019_PRX,Song_2019_PRL,Po_2019_PRB,bouhon2019wilson,Wang_2019_PRL} and $\mathbb{Z}_2$ rotation-winding number \cite{Chiu_2013_PRB,Zhang_2013_PRL,Chiu_2016_RMP}, protected by space time-reversal symmetry [$(C_{2z}\mathcal{T})^2=1$] and rotation symmetry $C_{2x}$, respectively. The combined symmetry $(C_{2z}\mathcal{T})^2=1$ imposes the reality condition on the Hamiltonian, leading to the real-valued corner state (Fig.\,\ref{fig:hoti}\textbf{b}). The $C_{2x}$ rotation-resolved Zak phase $\nu_{\pm}$ along the rotation-invariant line $k_y=0$, where $\pm$ denotes the rotation eigenvalue $c_{2x} = \pm1$, gives rise to a nontrivial $\mathbb{Z}_2$ rotation-winding number (Fig.\,\ref{fig:hoti}\textbf{c}).

We suggest a HOTI marker given by $\rchi^{\pm}(\mathbf{r}) = - \bra{\mathbf{r}} \widetilde{C}_{2x}^{\pm} P^{\pm}\hat{X}Q^{\pm} \ket{\mathbf{r}}$, where $\widetilde{C}_{2x}^{\pm} = P^{\pm} C_{2x} P^{\pm}$ is a projected symmetry operator, and $\hat{X}$ is a position operator (see Methods). Here, $P^{\pm}~(Q^{\pm})$ is the projection operator to the occupied (unoccupied) $c_{2x}=\pm1$ subspaces. In OBC, $\rchi^{\pm}(\mathbf{r})$ successfully diagnoses the rotation-winding number in real space: $\rchi^{\pm}(\mathbf{r})$ dictates the nontrivial rotation-winding number by being localized along the edge of the flake (Fig.\,\ref{fig:hoti}\textbf{d}), whereas in the trivial case, it is delocalized over the entire geometry (see Supplementary Figure 3). Interestingly, the corner state appears at the boundary between the opposite signs of each HOTI marker $\rchi^{\pm}(\mathbf{r})$. The sum of the opposite HOTI markers $\rchi^{+}(\mathbf{r}) + \rchi^{-}(\mathbf{r})$ is zero, which indicate a trivial winding number. The HOTI marker can be applied to symmetry-breaking perturbations, as demonstrated in TBG under the uniform magnetic field. 

To study the effect of the magnetic field on the HOTI states, we track the corner states by investigating their spectral flow at \textit{fixed filling}~\cite{herzog2020hofstadter} (see red and green lines in Fig.\,\ref{fig:energy}\textbf{b}). At zero flux, the highest occupied (HO) and lowest unoccupied (LU) states are identified as corner-localized states (Fig.\,\ref{fig:hoti}\textbf{b}).  They adiabatically evolve as a flux function and undergo a series of discontinuity transitions at specific fluxes. This discontinuity is indicative of a topological change due to bulk gap change \cite{herzog2020hofstadter,Lian20p041402}. Indeed, we reveal that HO and LU states at the discontinuity transitions are quantum Hall chiral edge states (see Supplementary Note 4).  

\begin{figure}[h]
\includegraphics[width=\textwidth]{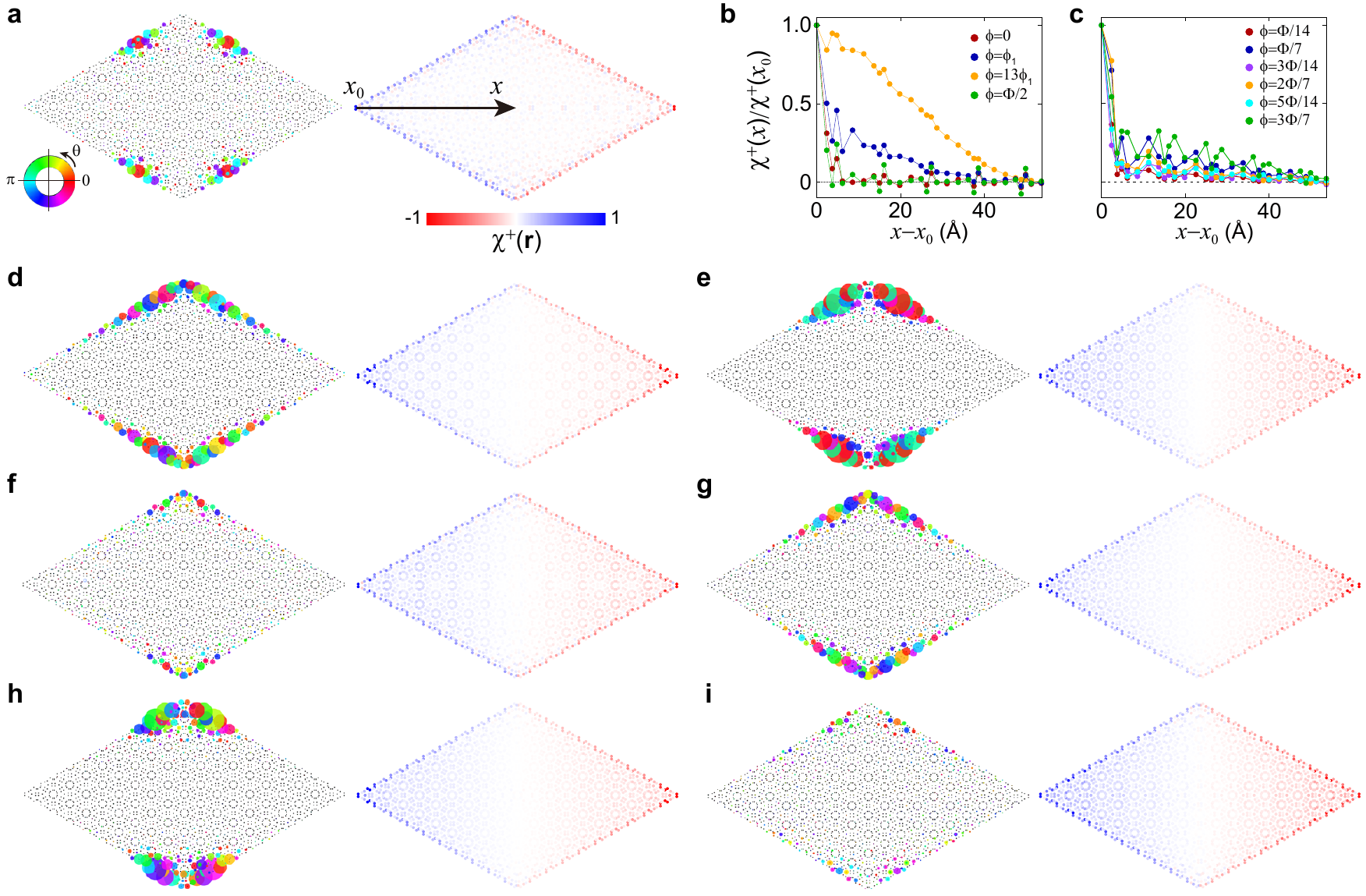}
\caption{
\label{fig:marker}
Characterization of replica HOTI states. \textbf{a,d-i} Real-space distribution of the corner states (the highest occupied eigenstates) and HOTI markers $\rchi^{+}(\mathbf{r})$ for various fluxes \textbf{a} $\phi=\frac{1}{2}\Phi$, \textbf{d} $\frac{1}{14}\Phi$, \textbf{e} $\frac{1}{7}\Phi$, \textbf{f} $\frac{3}{14}\Phi$, \textbf{g} $\frac{2}{7}\Phi$, \textbf{h} $\frac{5}{14}\Phi$, and \textbf{i} $\frac{3}{7}\Phi$.
In \textbf{a}, the colored circle indicates the phases of eigenstate components.
Due to the mirror symmetry about the flux $\phi=\Phi/2$, the eigenstates and markers at $\phi=\frac{p}{14}\Phi~ (p=1,2,\dots,6)$ are the same as those at $\phi=\frac{14-p}{14}\Phi$.
\textbf{b-c} Line profiles of $\rchi^{+}(\mathbf{r})$ along the arrow indicated in \textbf{a}.
Here, $\phi_1=\frac{1}{21000}\Phi$ and $x_0$ is the corner position.
}
\end{figure}

Remarkably, we find a reentrance of the HOTI phase at $\phi = \frac{1}{2}\Phi$, characterized by edge-localized marker $\rchi^{+}(\mathbf{r})$ (Fig.\,\ref{fig:marker}\textbf{a}).  $\rchi^{+}(\mathbf{r})$ decays exponentially along the bulk as $\exp[-\alpha (x-x_0)]$ with $\alpha=0.52$, which is identical to that of the exact HOTI state at zero flux (Fig.\,\ref{fig:marker}\textbf{b}) (see also Supplementary Note 3 for the detailed quantitative analysis). The re-entrant HOTI phase relies on composite symmetry $\mathcal{U}C_{2x}$ exactly preserved at $\phi = \frac{1}{2}\Phi$ because $(\mathcal{U}C_{2x})H(\frac{1}{2}\Phi)(\mathcal{U}C_{2x})^{\dagger}=H(\frac{1}{2}\Phi)$ from $C_{2x}H(\frac{1}{2}\Phi)C_{2x}^{\dagger}=H(-\frac{1}{2}\Phi)$ and $H(\phi+\Phi)=\mathcal{U}H(\phi)\mathcal{U}^\dagger$.  Note that the corner boundary modes of the re-entrant HOTI phase are localized at the corner, but the node appears slightly more concentrated off the corner (Fig.\,\ref{fig:marker}\textbf{a}) (see also Supplementary Figure 4 for the reason of the nodal structure of the corner states).

\subsection{Replica HOTIs}
In addition to the exact HOTIs at TRIFs ($\phi = 0$ and $\phi=\frac{1}{2}\Phi$), replicas of the original HOTIs are found at the $\frac{1}{7}\Phi$ quasi-periodic counterparts of TRIFs.   We employ HO and LU states as indicators of a replica of the original HOTI.  We find that they are positioned within the spectral gap at the specific fluxes of quasi-periodicity $\phi=\frac{p}{14}\Phi ~(p\in\mathbb{Z}; p\neq7\mathbb{Z}$) (Figs.\,\ref{fig:energy}\textbf{e-j}). A close inspection reveals that HO and LU states show oscillatory behavior of HO and LU energies as a function of flux, which originates from the Aharonov-Bohm tunneling in the presence of an external magnetic flux. Notably, the oscillation is a finite-size effect rather than a characteristic behavior of corner states, as is evident in the oscillations of other states near HO and LU states.  

To demonstrate the characteristics of the replica HOTIs, we plot the HO states in the left panels in Figs.\,\ref{fig:marker}\textbf{d-i}. The real-space distribution arguably shows the corner-localized states, supporting the HOTI phases. Nonetheless, these states exhibit stark contrast to the corner states of the exact HOTI at zero flux in that they are complex-valued functions, while the exact HOTI hosts real-valued corner states (Fig.\,\ref{fig:hoti}\textbf{b}). The complex-valued wave functions manifest the broken reality condition $[(C_{2z}\mathcal{T})^2=1]$ at finite fluxes, implying that the Stiefel-Whitney characterization is inapplicable. Furthermore, these quasiperiodic fluxes also break the $C_{2x}$ and $\mathcal{U}C_{2x}$ symmetries, which were utilized to characterize the exact HOTIs at TRIFs.

Remarkably, the HOTI marker can be defined without the protecting symmetries, enabling the evaluation of rotation-winding numbers. We find that the HOTI marker can quantitatively characterize the corner states in the presence of flux, that is, under rotational-symmetry $C_{2x}$ breaking. At a small flux $\phi_1=\frac{1}{21000}\Phi$, the eigenstate shows the remaining localized corner state, and the corner state is characterized by the marker $\rchi^{+}(\mathbf{r})$ which is sufficiently localized along the entire edge despite the small permeated values towards the bulk (Supplementary Figure 3). Quantitatively, $\rchi^{+}(\mathbf{r})$ exhibits an exponential decay as $\exp[-\alpha (x-x_0)]$ with $\alpha = 0.10$, which is smaller than $\alpha = 0.52$ of the exact HOTIs due to the symmetry breaking (Fig.\,\ref{fig:marker}\textbf{b}). The exponential localization of $\rchi^{+}(\mathbf{r})$ from the edge for the corner states is in stark contrast to a linear delocalization $\propto -\beta  (x-x_0)$ of $\rchi^{+}(\mathbf{r})$ along the whole geometry for the trivial state that occurs at, for example, $\phi=13 \phi_1$ (Fig.\,\ref{fig:marker}\textbf{b}). Such localization characteristics of the markers serve as a hallmark to identify nontrivial bulk topology, which fundamentally originates from the action of the projected symmetry operator, as in the generic topological crystalline insulating phases protected by spatial symmetries~\cite{mondragon2019robust,varjas_2020_computation} (see also Methods for the detailed explanation for the real-space behavior of the HOTI marker).

Our HOTI marker captures the replica HOTI phases as well, at $\phi=\frac{p}{14}\Phi ~(p\in\mathbb{Z}; p\neq7\mathbb{Z}$) (Figs.\,\ref{fig:marker}\textbf{d-i}). 
$\rchi^{+}(\mathbf{r})$ at $\phi=\frac{p}{14}\Phi$ show robust edge localization, consistent with the corner-localized eigenstates.
The line profiles of $\rchi^{+}(\mathbf{r})$ (Fig.\,\ref{fig:marker}\textbf{c}) exhibit exponential decay  (see also Supplementary Figure 6).
The replica HOTIs can be viewed as the copies of exact HOTIs disordered by the fractional flux quantum acquired when electrons travel through the minimal Peierls path because the composite symmetry $\mathcal{U}_0C_{2x}$ becomes exact when the interlayer coupling is turned off.

\begin{figure}[h]
\centering
\includegraphics[width=1\textwidth]{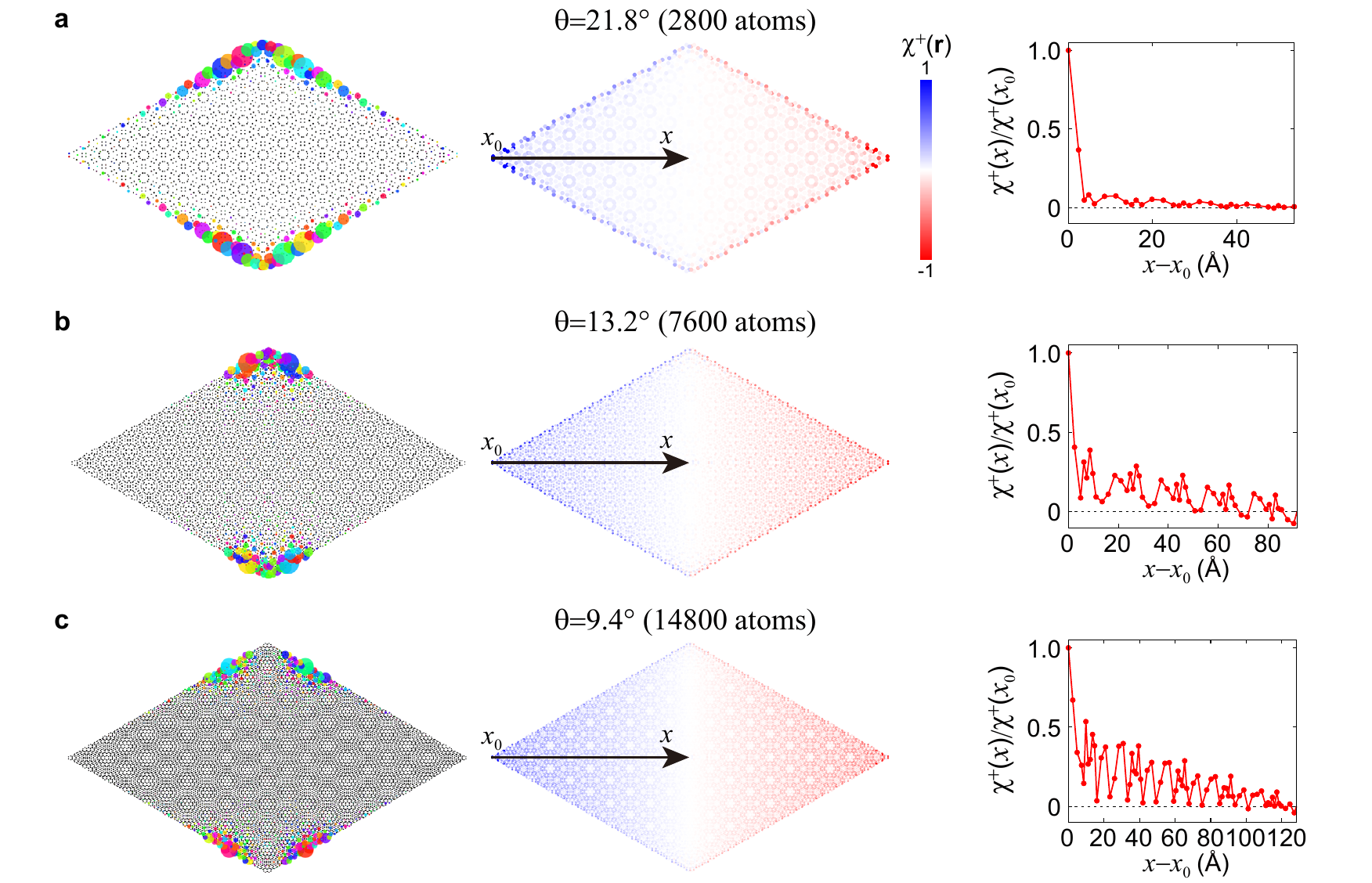}
\caption{
Replica HOTI states at other large angles. \textbf{a-c} Real-space distribution of corner states (the highest occupied eigenstates), HOTI marker $\rchi^{+}(\mathbf{r})$ and its line profile for the twist angles \textbf{a} $\theta_{1,2}=21.8^\circ$, \textbf{b} $\theta_{2,3}=13.2^\circ$, and \textbf{c} $\theta_{3,4}=9.4^\circ$. 
For the geometry in the open boundary condition, the $10\times10$ unit cell is used for the angles $\theta=21.8^\circ, 13.2^\circ$, and $9.4^\circ$, which contains 2800, 7600, 14800 atoms, respectively.
Here, we represent the replica HOTIs at \textbf{a} $\phi=\frac{1}{14}\Phi_{1,2}$, \textbf{b} $\phi=\frac{1}{38}\Phi_{2,3}$, and \textbf{c} $\phi=\frac{1}{74}\Phi_{3,4}$ where $\Phi_{m,n}$ is the flux periodicity at given integers $m$ and $n$.
}
\label{fig:rHOTI_angle}
\end{figure}

We also verify that replica HOTIs generally appear at other large angles. Figure \ref{fig:rHOTI_angle} shows the HO states and HOTI markers at the other twist angles $\theta=13.2^\circ ~(m=2,n=3)$ and $9.4^\circ ~(m=3,n=4)$. We find that both the corner localized states in real space and the localization behavior of the calculated HOTI markers support the existence of the replica HOTI states at the flux $\phi=\frac{p}{2N_{\textrm{rep}}}\Phi_{m,n} ~(p\in\mathbb{Z}; p\neq N_{\textrm{rep}}\mathbb{Z})$ where the flux periodicity is given by $\Phi_{m,n}=N_{\textrm{rep}}\Phi_{\textrm{G}}$ with the flux periodicity of graphene $\Phi_\textrm{G}$. 
Here, $N_{\textrm{rep}}=19$ and $37$ for $\theta=13.2^\circ$ and $9.4^\circ$, respectively. We note that the localization strength of the HOTI markers (see the line profiles in Fig.\,\ref{fig:rHOTI_angle}) is weakened as we decrease the twist angle because the bulk gap is significantly reduced (see Supplementary Figure 9).

We find that the out-of-plane rotational symmetry $C_{2x}$ is essential to realize the re-entrant exact and replica HOTI phases in TBG under a magnetic field. In contrast to our model, there is no re-entrant corner state at half-flux periodicity in the magic-angle TBG model \cite{herzog2020hofstadter} with only $C_{2z}\mathcal{T}$ symmetry, where the flux pumps corner states into the bulk.
The disappearance of the corner states at half periodicity confirms the inapplicability of the Stiefel-Whitney characterization for the HOTI states in the presence of a magnetic field. This indicates that additional crystalline symmetry, such as $C_{2x}$, is required to protect the corner states in TBG under a strong magnetic field.

In summary, we have demonstrated that HOTIs can occur without explicit protecting symmetries because of the self-similarity of Hofstadter butterflies as replicas of original HOTIs. We expect the distinct symmetry dependence of replica HOTIs can lead to distinct physical properties from the exact HOTIs (see Supplementary Note 6 for the detailed discussion). The HOTI marker is an invaluable tool for studying HOTI states in various situations beyond conventional methods using periodic boundary conditions. 
It offers the distinct advantage of being able to readily identify the HOTI phase, even at a small magnetic field in the open boundary condition. This is particularly advantageous compared to momentum-space methods relying on periodic boundary conditions, as they are computationally demanding at low magnetic fields, with their computational cost scaling inversely with the strength of the magnetic field.
The exponents of our HOTI marker allows for quantitative analysis, which can be potentially useful for future study such as many-body disordered HOTIs. The observation of the replica HOTI at the fixed filling requires a huge magnetic field B $\sim 10^5 \textrm{T}$, but replica topology may occur at different filling near low fields. Therefore, establishing an exact relationship between discrete scale invariance and band topology in this quantum fractal will be exciting future research with direct experimental implications. Additionally, a critical challenge that needs to be tackled in order to realize the observation is ensuring the stability of large TBG flakes under high magnetic fields. It would also be interesting to explore the Coulomb repulsion effect on replica phases at smaller angles, where the role of Coulomb repulsion is crucial~\cite{Kang_strong_2019,Vafek_renormalization_2020,Bernevig_interacting_2021,Lian_exact_2021,Bernevig_exact_2021,Song_magic_2022}. With much progress in synthesis of moir\'e materials ~\cite{doi:10.1126/sciadv.aay8409,Mre_ca_Kolasi_ska_2022,Park2021,Park2022,Burg2022,uri2023superconductivity} and measurement of Hofstadter energy spectrum~\cite{das2022observation,Yu2022_correlated}, our results can pave the way for studying replica topology under magnetic field in generic moir\'e multilayer~\cite{Park2021,Park2022,Burg2022} and moir\'e quasiperodic~\cite{uri2023superconductivity} systems that host multiple interaction scales.

\section{Methods}
\subsection{Tight-binding model}
We employ the Moon-Koshino tight-binding model for twisted bilayer graphene in Ref.\,\cite{Moon12p195458}, which is written as
\begin{eqnarray}
H=\sum_{ij} t_{ij}(\mathbf{R}_i-\mathbf{R}_j) c^\dagger_{\mathbf{R}_i}c_{\mathbf{R}_j} + h.c.,
\end{eqnarray}
where $c^\dagger_{\mathbf{R}_i}$ ($c_{\mathbf{R}_j}$) is a creation (annihilation) operator of an electron at the lattice site $\mathbf{R}_i$, and $t_{ij}(\mathbf{R}_i-\mathbf{R}_j)$ is the hopping integral between the sites $\mathbf{R}_i$ and $\mathbf{R}_j$.
The hopping integral is given by 
\begin{eqnarray}
-t_{ij}(\mathbf{R}_i-\mathbf{R}_j) = V_{pp\pi}\left[1-\left(\frac{\mathbf{d}\cdot\mathbf{\hat{z}}}{d}\right)^2\right]
+ V_{pp\sigma}\left(\frac{\mathbf{d}\cdot\mathbf{\hat{z}}}{d}\right)^2.
\end{eqnarray}
Here, the hopping parameters are given as a decaying function of a hopping distance $d = |\mathbf {d}| = |\mathbf{R}_i-\mathbf{R}_j|$
\begin{eqnarray}
V_{pp\pi}=V_{pp\pi}^0\exp\left(-\frac{d-a_0}{\delta}\right)~~\text{and}~~
V_{pp\sigma}=V_{pp\sigma}^0\exp\left(-\frac{d-d_0}{\delta}\right),
\end{eqnarray}
where $a_0\approx1.42\text{\AA}$ is the bond length of graphene, $d_0\approx3.35\text{\AA}$ is the interlayer distance, and $\delta=0.319a_0$ is the decay length.
Here, we set $V_{pp\pi}^0=-2.7$ eV and $V_{pp\sigma}^0=0.82$ eV, which reproduce the band structure of 21.8$^{\circ}$ twisted bilayer graphene with a bulk gap $\sim 9$ meV in a HOTI state\,\cite{MJPark1,MJPark2}.
Our tight-binding model under a periodic boundary condition (see atomic geometry used in Supplementary Figure 1) has 14 occupied $p_z$ orbital bands that consist of the same number of $c_{2x}=+1$ and $c_{2x}=-1$ bands, where $c_{2x}$ is an eigenvalue of a twofold rotational symmetry operator $C_{2x}$ about the $x$-axis. 
This implementation of the model successfully reproduces the nontrivial rotation-winding number (Fig. \ref{fig:hoti}\textbf{c}) in line with the previous DFT results\,\cite{MJPark1}. 
For the calculations of Hofstadter butterflies and topological markers, we use the flake geometry with an open boundary condition (see Supplementary Figure 1).

To study the effect of the magnetic field, we incorporate a magnetic flux $\phi$ into the hoppings as an additional phase via Peierls substitution\,\cite{peierls1933theorie}: 
\begin{eqnarray}
t_{ij}(\mathbf{R}_i-\mathbf{R}_j)
&\rightarrow& t_{ij}(\mathbf{R}_i-\mathbf{R}_j)\exp{\left(i\frac{e}{\hbar}\int_{\mathbf{R}_j}^{\mathbf{R}_i} \mathbf{A}_0\cdot d\mathbf{r}\right)} \nonumber
\\
&=& 
t_{ij}(\mathbf{R}_i-\mathbf{R}_j)\exp{\left[i\frac{e}{\hbar}\frac{\phi}{2S_{\mathrm{min}}}(x_i+x_j)(y_j-y_i)\right]},
\end{eqnarray}
where the vector potential $\mathbf{A}_0=B(0,x)=\frac{\phi}{S_{\mathrm{min}}}(0,x)$ for $\mathbf{B}=B\hat{z}$ and $S_{\mathrm{min}}$ is the interior area of the minimal Peierls path.
We prove that our twisted bilayer graphene lattice has exact flux periodicity (see Supplementary Note 2).
The Hofstadter energy spectrum of our system is thus periodic under the translation by a magnetic flux quantum $\Phi = \frac{h}{e}$ because the Hamiltonian can be gauge transformed according to\,\cite{herzog2020hofstadter}  
\begin{eqnarray}
H(\phi+\Phi)=\mathcal{U}(\mathbf{A})H(\phi)\mathcal{U}^\dagger(\mathbf{A}).
\end{eqnarray}
The unitary matrix $\mathcal{U}(\mathbf A)=\sum_{\mathbf{R}} c^\dagger_{\mathbf{R}}c_{\mathbf{R}}\exp{(i\frac{e}{\hbar}\int_{\mathbf{r}_0}^{\mathbf{R}}{\mathbf {A}}\cdot d\mathbf{r})}$ ($\mathbf{r}_0$: a fixed lattice site) is defined for the vector potential $\mathbf{A} (\neq \mathbf{A}_0)$ that leads to the flux quantum $\int_{S_{\mathrm{min}}}(\nabla\times{\mathbf{A}}) \cdot d\mathbf{S}=\Phi$.

\subsection{Kernel polynomial method}
Hofstadter butterflies of twisted bilayer graphene can be efficiently calculated by using the kernel polynomial method \cite{RevModPhys.78.275}. The essential idea of the kernel polynomial methods is to expand the density of states $\rho(E)$ ($E$: energy) in terms of Chebyshev polynomials as,
\begin{eqnarray}
\label{eq_KPM_dos}	
	\rho (E) = \pi \sqrt{1-E^2} \sum_{n=0}^{M} \mu_{n} U_{n}(E),  
\end{eqnarray}
where
$U_{n}(E)$ is the second kind $n$-th Chebyshev polynomials, 
\begin{eqnarray}
\label{eq_KPM_poly}	
	U_{n}(E)= \frac{\mathrm {sin}[(n+1) \mathrm {arccos}(E)]}{\mathrm{sin}[\mathrm {arccos}(E)]}.  
\end{eqnarray}
Here, $\mu_{n}$ is the moment for an operator $\hat{O}(E)$, which reads
\begin{eqnarray}
\label{eq_KPM_moments}
	\mu_{n} = \frac{2}{\pi^2} \int_{-1}^{1} dE \,  \hat {O} (E) U_n(E).
\end{eqnarray}	
The targeting density of states operator $\hat{\rho} (E)$ is given by
\begin{eqnarray} 
\label{eq_KPM_op_dos}
	\hat{\rho} (E) = \frac {1}{N} \sum_{k=1}^N \delta (E-E_k).
\end{eqnarray}
After putting $\hat{\rho} (E)$ into $\mu_{n}$, we obtain
\begin{eqnarray} 
\label{eq_KPM_moment_tr}
	\mu_n &=&\frac{2}{\pi^2}  \int_{-1}^{1} dE \,  \hat {\rho} (E) U_n(E) \nonumber \\
				&=&\frac{2}{\pi^2} \frac{1}{N} \sum_{k=1}^{N} U_n(E_k)  \nonumber  \\
				&=&\frac{2}{\pi^2} \frac{1}{N} \sum_{k=1}^{N} \left\langle k \left |U_n(H) \right | k  \right\rangle \nonumber \\
				&=&\frac{2}{\pi^2} \mathrm {Tr} (U_n(H)).
\end{eqnarray}	
A stochastic approach is employed to obtain the trace by introducing the $R$-number of random vectors $|r\rangle$, instead of (potentially unknown) exact eigenvectors:
\begin{eqnarray} 
\label{eq_KPM_tr_u}
	\mathrm{Tr} (U_m (H)) \simeq \frac{1}{R} \sum_{r=1}^{R} \left\langle r \left | U_m (H) \right | r \right\rangle ,
\end{eqnarray}	
where $R$ is set to a sufficiently large value to attain the converged density of states.
Then, we take advantage of a recursive relation for the polynomial,
\begin{eqnarray} 
\label{eq_KPM_rec_u}
	U_{m+1}=& 2H U_m -U_{m-1},
\end{eqnarray}	
to rewrite the trace as
\begin{eqnarray} 
\label{eq_KPM_tr_rand}
	\mathrm {Tr} (U_m(H)) \simeq \frac{1}{R} \sum_{r=1}^R \langle r | r \rangle_m,
\end{eqnarray}	
where 
\begin{eqnarray} 
\label{eq_KPM_r_rec}
	| r \rangle_{m} =& U_m (H) 	| r \rangle ~~\text{and}~~
	| r \rangle_{m+1} =& 2H | r \rangle_{m} - | r \rangle_{m-1}.
\end{eqnarray}	
As a result, the density of states is obtained as 
\begin{eqnarray}
\label{eq_KPM_res_dos}	
	\rho (E) = \frac{2}{\pi} \sqrt{1-E^2} \sum_{n=0}^{M} g_m^M \mathrm{Tr}(U_m(H)) U_{n}(E),  
\end{eqnarray}
where $g_m^M$ is the Jackson kernel,
\begin{eqnarray}
\label{eq_KPM_kernel}	
	g_m^M =  \frac{1}{M+1} \left[ (M-m+1) \mathrm {cos}\frac{m\pi}{M+1} + \mathrm {sin}\frac{m\pi}{M+1}\mathrm {cot} \frac{\pi}{M+1} \right ],
\end{eqnarray}	
which is introduced to reduce the Gibbs oscillation \cite{RevModPhys.78.275}.

\subsection{HOTI topological marker}
Topological marker is a local quantity in real space that characterizes the topological phases~\cite{bianco2011mapping,Shem2014topological,tran2015topological,caio2019topological,mondragon2019robust}.
The local Chern marker was first introduced as a topological marker whose spatial average in bulk in thermodynamic limit corresponds to the Chern number of the system~\cite{bianco2011mapping}.
The topological marker was then generalized to the topological crystalline insulating (TCI) phases, in which the topological states are protected by the spatial symmetries~\cite{mondragon2019robust}.
The generalized topological marker $\mathcal{T}_G(\mathbf{r})$ related to the symmetry $G$ is given by
\begin{eqnarray}
\mathcal{T}_G(\mathbf{r})=\bra{\mathbf{r}}\widetilde{G}\mathcal{F}(P)\ket{\mathbf{r}}, \label{eq:top_marker_G}
\end{eqnarray}
where $\widetilde{G} = PGP$ is a projected symmetry operator and a function $\mathcal{F}(P)$ encodes the types of topological invariants.
For example, $\mathcal{F}(P)\propto P[\hat{X},P]$ and $P[[\hat{X},P],[\hat{Y},P]]$ for 1D winding~\cite{Shem2014topological} and 2D Chern numbers~\cite{bianco2011mapping}, respectively, where $\hat{X}$ and $\hat{Y}$ are position operators. 

We extend the topological marker to a HOTI version in our twisted bilayer graphene system.
The extension is straightforward because the HOTI phase in twisted bilayer graphene is protected by the $C_{2x}$ rotation symmetry resolved winding number, the rotation-winding number.
Let us first see the topological marker $\rchi(\mathbf{r})$ for the $C_{2x}$ symmetry, which is given by
\begin{eqnarray}
\rchi(\mathbf{r}) \equiv  \mathcal{T}_{C_{2x}}(\mathbf{r}) = \bra{\mathbf{r}}\widetilde{C}_{2x}P[\hat{X},P]\ket{\mathbf{r}}
=-\bra{\mathbf{r}}\widetilde{C}_{2x}P\hat{X}Q\ket{\mathbf{r}},
\end{eqnarray}
where we used the relation $Q=1-P$ in the last equality. 
By projecting the projection operators to the $C_{2x}$ rotation $\pm$ subspaces as $P=P^+ + P^-$ and $Q=Q^+ + Q^-$, we obtain
\begin{eqnarray}
\rchi(\mathbf{r})
&=&-\bra{\mathbf{r}}\widetilde{C}_{2x}P\hat{X}Q\ket{\mathbf{r}} \nonumber
\\
&=& -\bra{\mathbf{r}}(P^+C_{2x}P^+ + P^-C_{2x}P^-)(P^+\hat{X}Q^+ + P^-\hat{X}Q^-)\ket{\mathbf{r}}  \nonumber
\\
&=& -\bra{\mathbf{r}}(P^+C_{2x}P^+\hat{X}Q^+ + P^-C_{2x}P^-\hat{X}Q^-)\ket{\mathbf{r}} \label{eq:marker}
\\
&\equiv& \rchi^+(\mathbf{r}) + \rchi^-(\mathbf{r}),
\nonumber 
\end{eqnarray}
where we used the condition $P^+P^-=Q^+Q^-=PQ=0$.
The $C_{2x}$ rotation-resolved topological marker $\rchi^{\pm}(\mathbf{r})$ serves as the real space local expression of the rotation-resolved Zak phase $\nu_{\pm}$.

\subsection{Real-space behavior of HOTI marker}

To understand the real-space behavior of the HOTI marker, we first consider the localization property of the topological markers for TCI phases. The topological markers for TCI phases feature the exponential localization from the subspace restricted by the spatial symmetries~\cite{mondragon2019robust,varjas_2020_computation}. 
It is different from the case of the typical Chern insulators without symmetries where the localization sites of the topological marker are all the sites within the bulk~\cite{bianco2011mapping,tran2015topological,caio2019topological}.
In a TCI phase, protected by a spatial symmetry $G$, the eigenvalues of $G$ classify the eigenstates of the Hamiltonian $\{\ket{u_i}\}$ and thus the projection matrix for occupied states $P=\sum_{n\in occ.}\ket{u_n}\bra{u_n}$ at the symmetry-invariant subspace $\mathcal{S}$.
As a result, the bulk topology of a TCI phase is encoded by the projection matrix $P$ at the invariant subspace $\mathcal{S}$, which allows the introduction of the real-space topological invariant, the topological marker $\mathcal{T}_G(\mathbf{r})$ in Eq.~\ref{eq:top_marker_G}.
It is proven that the topological marker $\mathcal{T}_G(\textbf{r})$ exhibits the exponential localization from the fixed points $\textbf{r}_S(\in\mathcal{S};\mathcal{S}=\{\textbf{r}|G\textbf{r}=\textbf{r}\})$ of the spatial symmetry $G$ as~\cite{mondragon2019robust}
\begin{eqnarray}
|\mathcal{T}_G(\textbf{r})|< \mathcal{O}(e^{-|\textbf{r}_S-\textbf{r}|/\zeta}) ~~~~ \textrm{when} ~~~~ |\textbf{r}_S-\textbf{r}|\gg\zeta. 
\end{eqnarray}
Here the length scale $\zeta$ is rough in the order of the inverse gap/localization strength. We note that the localization property is fundamentally arising from the action of the projected symmetry operator $\widetilde{G}$: the projection matrix $P$ is exponentially localized for the insulators~\cite{Kohn_1996_Density,resta2006kohn,resta2011insulating} and the symmetry $G$ restricts the localization site of the markers~\cite{mondragon2019robust,varjas_2020_computation}.
The localization property is more general than the exponentially localized Wannier functions because the topological marker is localized even in the presence of a nonzero Chern number, which prohibits the construction of localized Wannier functions.

In the case of $C_{2x}$ winding number, the presence of the projected symmetry operator $\widetilde{C}_{2x}^{\pm}$ gives rise to the exponential localization of the HOTI marker $\rchi^{\pm}(\mathbf{r})$ from the edge.
The absence of the winding number allows for the  $C_{2x}$ specification of the winding number, revealing the presence of the $C_{2x}$-protected metallic edge states. As in the case of the known HOTI phases characterized by $C_{2x}$ winding number~\cite{MJPark1,MJPark2}, the corner state is the Su-Schrieffer-Heeger type domain wall state, arising from the gap opening of the edge states. 
The edge is the bulk of the corner states in the HOTI state. Due to the action of the projected symmetry operator $\widetilde{C}_{2x}^{\pm}$, our HOTI marker of HOTI states exhibits exponential localization from the edge $\textbf{r}_\textrm{E}$ as 
\begin{eqnarray}
|\rchi^{\pm}(\mathbf{r})|< \mathcal{O}(e^{-|\textbf{r}_\textrm{E}-\textbf{r}|/\zeta}) ~~~~ \textrm{when} ~~~~ |\textbf{r}_\textrm{E}-\textbf{r}|\gg\zeta. 
\end{eqnarray}
On the contrary, HOTI trivial cases do not show such localization behavior from the edge.
Instead, they are linearly delocalized over the entire geometry ($\propto$ the position operator $\hat{X}$) which follows from the form of the $C_{2x}$-marker formula proportional to $\hat{X}$ in Eq.~\ref{eq:marker}. 

\section{Data availability}
The authors declare that the data supporting the findings of this study are available within the article and its supplementary information files or from the corresponding authors on reasonable request.

\section{Code availability}
The code generated during this study is available from the corresponding author upon reasonable request.

\section{Acknowledgements}
\begin{acknowledgments}
S.-W.K. thanks Jonah Herzog-Arbeitman for helpful discussions. M.J.P. thanks Jaehoon Kim for providing mathematical insights. This work was supported by the Korean National Research Foundation (NRF) Basic Research Laboratory (NRF-2020R1A4A307970713), the NRF Grant numbers (NRF-2021R1A2C101387112 and NRF-2021M3H3A1038085). This work was also supported by the National Research Foundation of Korea (NRF) grant funded by the Korea government (MSIT) (RS-2023-00252085, RS-2023-00218998). The computational resource was provided by the Korea Institute of Science and Technology Information (KISTI) (KSC-2020-CRE-0108).
\end{acknowledgments}

\section{Author contributions}
Y.K. conceived the idea and organized the research. Y.K. and M.J.P. supervised the study. S.-W.K. and S.J. calculated the Hofstadter butterfly spectra. S.-W.K. developed the higher-order topological marker and analyzed the topological phases. M.J.P provided the mathematical proof of the exact flux translational symmetry. All authors discussed the results and contributed to writing the manuscript.

\section{Competing interests}
The authors declare no competing financial or non-financial interests.

\def\bibsection{

\ifarXiv
    

\section*{Supplementary material}

**Supplementary Note 1. Atomic structure of twisted bilayer graphene**

Supplementary Figure 1 shows atomic structures of 21.8° twisted bilayer graphene used for the calculations with periodic and open boundary conditions in the main text.

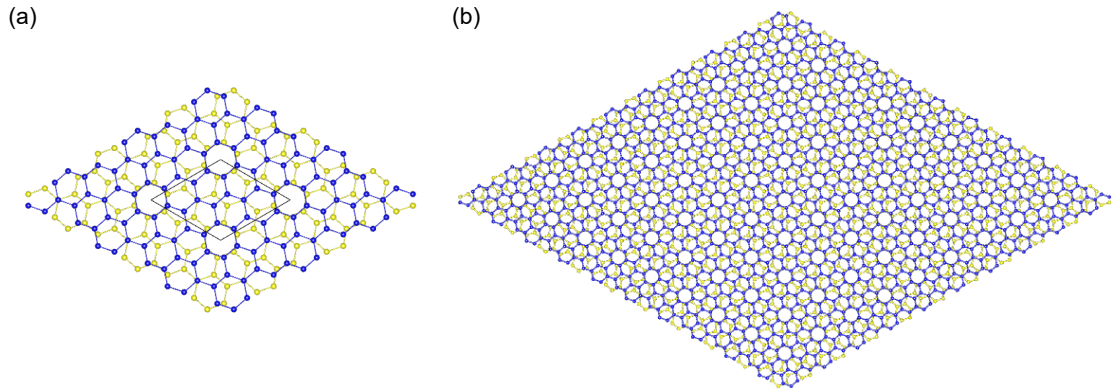

Supplementary Figure 1: Atomic structures of 21.8° twisted bilayer graphene used for the calculations with (a) periodic and (b) open boundary conditions. In (a), the black line indicates the $1 \times 1$ unit cell that contains 28 carbon atoms. In (b), rhombus flake geometry, whose size corresponds to the $10 \times 10$ unit cell area, contains 2800 carbon atoms.

**Supplementary Note 2. Exact flux periodicity of twisted bilayer graphene lattice**

In general lattice models with multiple sites per unit cell, the area of the two arbitrary Peierls loops may not be commensurate (i.e., the ratio of two areas is given by the irrational number). In such a case, there is no rigorous magnetic translational symmetry. In contrast, we show that in the case of the AA stacked twisted bilayer graphene, any arbitrary Peierls loops are commensurate with each other in all commensurate angles. Moreover, the area of all Peierls loops can be expressed as an integer multiple of the minimal Peierls loop.

Consider the set of commensurate twist angles, which can be generated by the two integers, $(m, n)$, as,

$$\cos\theta = \frac{1}{2}\frac{m^2 + n^2 + 4mn}{m^2 + n^2 + mn}, \quad \sin\theta = \frac{\sqrt{3}}{2}\frac{|m^2 - n^2|}{m^2 + n^2 + mn}, \tag{1}$$

where $0 < \theta < \pi/3$. The size of the moire unit cell is amplified by $N_L = m^2 + n^2 + mn$ times compared to the monolayer graphene.

**Claim 1**: In the twisted bilayer of the AA stacked honeycomb lattice, the minimal area of the triangular Peierls path is given as,

$$S_{\min}^{\text{TBG}} = \frac{\sqrt{3}}{4N_L}\gcd(z_1, z_2, z_3) \tag{2}$$

where $z_1 = (m^2 - n^2)$, $z_2 = (3m^2)$, $z_3 = (2m^2 - mn - n^2)$. gcd indicates the greatest common divisor. Here, we set the lattice constant as $a = 1$ for the simplicity. Moreover, the area of all possible triangular Peierls loops are divisible by $S_{\min}$.

*Proof*: We express the coordinates of the lattice sites on the unrotated layer(top layer) as,

$$\begin{aligned}
\mathbf{r}_t \equiv (x_t, y_t) &= (1, 0) + p\mathbf{a}_1 + q\mathbf{a}_2 + \delta(1, 0) \\
&= (1, 0) + p(\frac{3}{2}, \frac{\sqrt{3}}{2}) + q(0, \sqrt{3}) + \delta(1, 0), \\
&= (1 + \frac{3}{2}p + \delta, \frac{\sqrt{3}}{2}p + \sqrt{3}q), \tag{3}
\end{aligned}$$

where $p, q$ are arbitrary integers. $\delta = 0\ (1)$ is a constant for A (B) sublattice. Then, the set of the coordinates of the bottom layer can be written as, $\mathbf{r}_b = R(\theta)\mathbf{r}_t$, which is explicitly

given as,

$$\begin{aligned}
x_b &= \cos\theta x_t - \sin\theta y_t, \\
&= \frac{1}{2N_L}[(m^2 + n^2 + 4mn)x_t - \sqrt{3}|m^2 - n^2|y_t], \\
y_b &= \sin\theta x_t + \cos\theta y_t \\
&= \frac{1}{2N_L}[\sqrt{3}|m^2 - n^2|x_t + (m^2 + n^2 + 4mn)y_t].
\end{aligned} \quad (4)$$

To find the minimal area of the triangular Peierls path, without loss of generality, we only need to consider the triangles, where the two of the vertices consist of the top layer sites $[\mathbf{r}_{t,1} = (x_1, y_1), \mathbf{r}_{t,2} = (x_2, y_2)]$ and the other consists of the bottom layer site $[\mathbf{r}_b = (x_b, y_b) = R(\theta)(x_3, y_3)]$. The area of such triangles can be written as,

$$S = \frac{1}{2}|x_{t,1}(y_{t,2} - y_b) + x_{t,2}(y_b - y_{t,1}) + x_b(y_{t,1} - y_{t,2})| \quad (5)$$

We explicitly plug in Eq. (4). Since each $x, y$ coordinates are the linear polynomial of $(p_1, p_2, p_3, q_1, q_2, q_3) \in \mathbb{Z}^6$, the expression for the area of the triangles can be written in terms of the quadratic polynomials of $(p_1, p_2, p_3, q_1, q_2, q_3)$,

$$S = \frac{\sqrt{3}}{4N_L}\left|z_0 + \sum_i z_{p,i}p_i + \sum_i z_{q,i}q_i + \sum_{i \neq j} z_{pp,ij}p_ip_j + \sum_{i \neq j} z_{pq,ij}p_iq_j + \sum_{i \neq j} z_{qq,ij}q_iq_j\right| \quad (6)$$

where $z$ are the non-zero integer coefficients of the polynomial. There are total 31 $z$ coefficients and the explicit expression will be given later. To further simplify the above expression, we use the theorem of the linear Diophantine equation [1], which states as,

**Theorem 1** : The linear Diophantine equation, which is given as, $ax + by = c$, where $a, b, c$ are given integers, has a integer solution, $(x, y)$, if and only if $c$ is a multiple of gcd(a,b), where gcd is the greatest common divisor.

i.e. $\{c|c = ax + by, \ \forall a, b \in \mathbb{Z}\} = \{Nx|x = \gcd(a,b), \ N \in \mathbb{Z}\}$

By sequentially applying **Theorem 1** to Eq. (6), we find that the set of the areas of the possible triangles is given as,

$$A = \{S|\ S = \frac{\sqrt{3}}{4N_L}|z_0 + N\gcd(\mathbf{z}_p, \mathbf{z}_q, \mathbf{z}_{pp}, \mathbf{z}_{pq}, \mathbf{z}_{qq})|,\ N \in \mathbb{Z}\} \tag{7}$$

Here, the input of gcd includes all $z$ coefficients except for $z_0$. We omit the $i, j$ subscript for $z$ for the notational simplicity.

We now calculate $\gcd(\mathbf{z}_p, \mathbf{z}_q, \mathbf{z}_{pp}, \mathbf{z}_{pq}, \mathbf{z}_{qq})$. To do so, we can ignore the duplicating coefficients, $z_i$. In addition, if $z_i$ can be expressed as the linear integer polynomial of other coefficients $z_\alpha, z_\beta, ...z_\omega$. By **Theorem 1**, we can ignore $z_i$ since $\gcd(z_i, z_\alpha, z_\beta, ...z_\omega) = \gcd(N\gcd(z_\alpha, z_\beta, ...z_\omega), z_\alpha, z_\beta, ...z_\omega) = \gcd(z_\alpha, z_\beta, ...z_\omega)$. We divide into the three different cases depending on the sublattice degree of freedom.

(i) When all three vertices belong to the same sublattice $[(\delta_1, \delta_2, \delta_3) = (0, 0, 0)]$. The distinct coefficients, $z'$, together with $z_0$ are given as,

$$z_0 = 0,\ z'_1 = 2m^2 - mn - n^2,\ z'_2 = 3m^2 - 3n^2,\ z'_3 = 6mn + 3n^2. \tag{8}$$

Here, the subscripts of $z'$ are given in arbitrary order.

(ii) When one of the top layer vertices belong to the different sublattice $[(\delta_1, \delta_2, \delta_3) = (1, 0, 0)]$, and (iii) when the bottom layer belong to the different sublattice $[(\delta_1, \delta_2, \delta_3) = (1, 1, 0)]$, the distinct coefficients, $z''$, together with $z_0$ are given as,

$$z_0 = m^2 - n^2,\ z''_1 = 2m^2 - mn - n^2,\ z''_2 = 3m^2 - 3n^2,\ z''_3 = 3m^2, \tag{9}$$

Finally, we arrive at the identity, $\gcd(\mathbf{z}_p, \mathbf{z}_q, \mathbf{z}_{pp}, \mathbf{z}_{pq}, \mathbf{z}_{qq}) = \gcd(z'_1, z'_2, z'_3)$, and the expression for the area of the triangle is given as,

$$A = \{S|\ S = \frac{\sqrt{3}}{4N_L}|z_0 + N\gcd(z'_1, z'_2, z'_3)|,\ N \in \mathbb{Z}\}. \tag{10}$$

In case (i), $z_0 = 0$, the set of the area is explicitly given as,

$$A_{(i)} = \{NS|\ S = \frac{\sqrt{3}}{4N_L}\gcd(z'_1, z'_2, z'_3),\ N \in \mathbb{Z}^+\}. \tag{11}$$

All the possible area is an integer multiple of the minimal area, which is given as,

$$S_{\min,(i)} = \frac{\sqrt{3}}{4N_L} \gcd(2m^2 - mn - n^2, 3m^2 - 3n^2, 6mn + 3n^2) \tag{12}$$

In case (ii) and (iii), if $z_0 = m^2 - n^2$ is a multiple of $\gcd(z_1'', z_2'', z_3'')$, the set of the area is explicitly given as,

$$A_{(ii),(iii)} = \{NS|\ S = \frac{\sqrt{3}}{4N_L} \gcd(z_1'', z_2'', z_3''),\ N \in \mathbb{Z}^+\}. \tag{13}$$

if $z_0 = m^2 - n^2$ is not a multiple of $\gcd(z_1'', z_2'', z_3'')$, $\gcd(z_1'', z_2'', z_3'')$ must be $z_2'' = 3(m^2 - n^2)$. Then,

$$A_{(ii),(iii)} = \{(3N+1)S|\ S = \frac{\sqrt{3}}{4N_L}(m^2 - n^2),\ N \in \mathbb{Z}^+\}. \tag{14}$$

In both cases of (ii),(iii), all the possible area is an integer multiple of the minimal area, which is given as,

$$S_{\min,(ii),(iii)} = \frac{\sqrt{3}}{4N_L} \gcd(m^2 - n^2, 3m^2, 2m^2 - mn - n^2). \tag{15}$$

**Theorem 2** : $S_{\min,(i)}$ is divisible by $S_{\min,(ii),(iii)}$.

*Proof* : We can explicitly write down the linear equation such that,

$$z_3' = 6mn + 3n^2 = -6(2m^2 - mn - n^2) + 3(m^2 - n^2) + 3(3m^2) \tag{16}$$

Then by **Theorem 1**,

$$\begin{aligned}\gcd(z_1', z_2', z_3') &= \gcd(z_1', z_2', N\gcd(m^2 - n^2, 3m^2, 2m^2 - mn - n^2)) \\ &= \gcd(z_1', z_2', \gcd(N(m^2 - n^2), N(3m^2), N(2m^2 - mn - n^2))) \\ &= \gcd(3(m^2 - n^2), N(m^2 - n^2), N(3m^2), 2m^2 - mn - n^2),\end{aligned}$$

which is divisible by $\gcd(m^2 - n^2, 3m^2, 2m^2 - mn - n^2)$. As a result, $S_{\min,(i)}$ is divisible by $S_{\min,(ii),(iii)}$. ∎

Since $S_{\min} \equiv S_{\min,(ii),(iii)}$ divides $S_{\min,(i)}$, $S_{\min}$ divides every elements of $A_{(i)}$, $A_{(ii)}$, $A_{(iii)}$,

which concludes the proof of **Claim 1**. ∎

Finally, for the general Peierls loop visiting $n$ number of sites, the area of the Peierls loop can always be divided into multiple triangular loops (e.g., see Supplementary Figure 2), with either positive or negative contributions. As a result, the area of all $n$-site Peierls loop is the integer multiples of $S_{\min}^{\mathrm{TBG}}$.

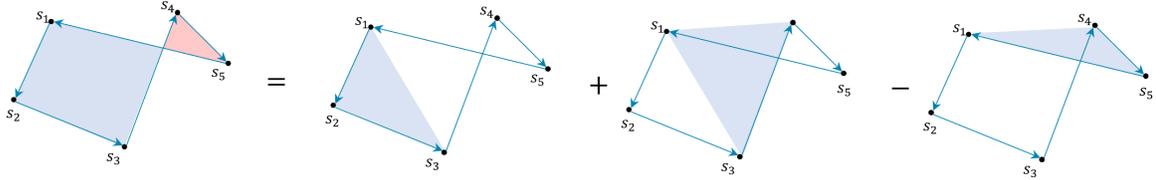

Supplementary Figure 2: Exemplary triangulation of the Peierls loop with five sites. The blue and the red regions indicate the positive and the negative flux area that the electron encloses, respectively. The Peierls loop can be divided into multiple triangular loops.

**Supplementary Note 3. Caluculated results using the HOTI marker**

*3.1. Exact HOTI states at $\phi = 0$ and $\frac{1}{2}\Phi$*

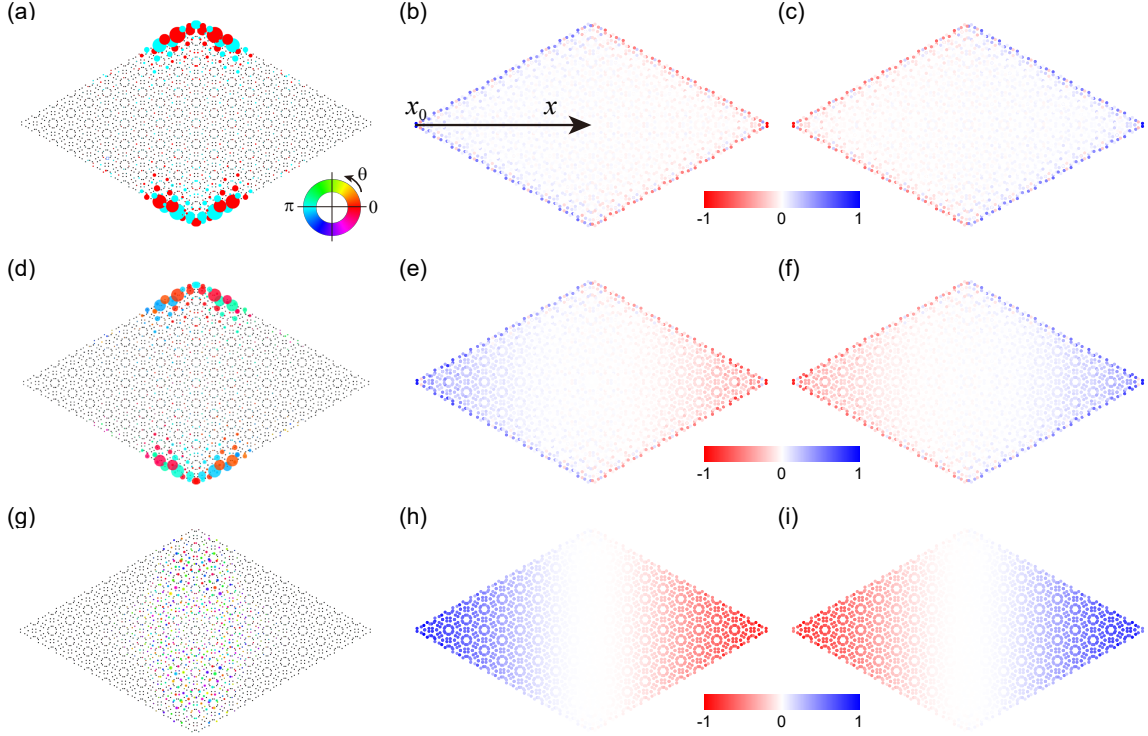

Supplementary Figure 3: (a)-(c) The highest occupied eigenstate, $\mathcal{X}^+(\mathbf{r})$, and $\mathcal{X}^-(\mathbf{r})$ of 21.8° twisted bilayer graphene in open boundary condition at zero flux. In (a), the colored circle indicates the phases of eigenstate components. (d)-(f) Same data set as (a)-(c) but at the flux $\phi_1 = \frac{1}{21000}\Phi$. (g)-(i) Same data set as (a)-(c) but at the flux $\phi = 13\phi_1$. For the markers, the color scales are normalized by the maximum value of $\mathcal{X}^+(\mathbf{r})$ at each flux.

At zero flux, the HOTI markers $\mathcal{X}^+(\mathbf{r})$ and $\mathcal{X}^-(\mathbf{r})$ exhibit opposite finite values localized at the edge of the flake [Supplementary Figures 3(b) and 3(c)] in which their sum $\mathcal{X}(\mathbf{r})$ is zero along the entire geometry. The corner state appears at the boundary between opposite signs of each HOTI marker [Supplementary Figure 3(a)]. This is consistent with the non-trivial rotation-winding number calculated in periodic boundary conditions (Fig. 3c in the main text). When the small flux $\phi_1 = \frac{1}{21000}\Phi$ is inserted, the eigenstate shows remaining localized corner state [Supplementary Figure 3(d)] and the corner state is characterized by the marker $\mathcal{X}^+(\mathbf{r})$ which is sufficiently localized along the entire edge despite the small permeated values towards the bulk [Supplementary Figures 3(e) and 3(f)]. If the flux further

8

increases to $\phi = 13\phi_1$, the corner state disappears and thus the eigenstate becomes a trivial state [Supplementary Figure 3(g)]. The marker is delocalized along the entire geometry [Supplementary Figures 3(h) and 3(i)], in stark contrast to the edge-localized marker of the HOTI state [Supplementary Figures 3(b) and 3(c)].

The exact HOTI phase re-enters at $\phi = \frac{1}{2}\Phi$, which is characterized by corner-localized boundary modes as well as the edge-localized marker (Fig. 4**a** in the main text). The re-entrant exact HOTI state is protected by the composite symmetry $\mathcal{U}C_{2x}$ that plays a role of $C_{2x}$ symmetry in the calculation of the marker in Eq. (24) in the main text. Note that the corner boundary modes of the re-entrant HOTI phase are localized at the corner, but the node appears slightly more concentrated off the corner [Supplementary Figure 4(a)]. The unusual shape of the corner states at the half-flux periodicity $\phi = \frac{1}{2}\Phi$ arises due to the $\mathcal{U}C_{2y}$ symmetry. The corner states are an odd function of $x$ as they respect $\mathcal{U}C_{2y}$ by having a definite eigenvalue $-1$ under the symmetry operation. This leads to the nodal structure at the $x = 0$ line in contrast to the nodal-free corner states at zero flux having $C_{2y}$ eigenvalue $+1$ [Supplementary Figure 4(b)].

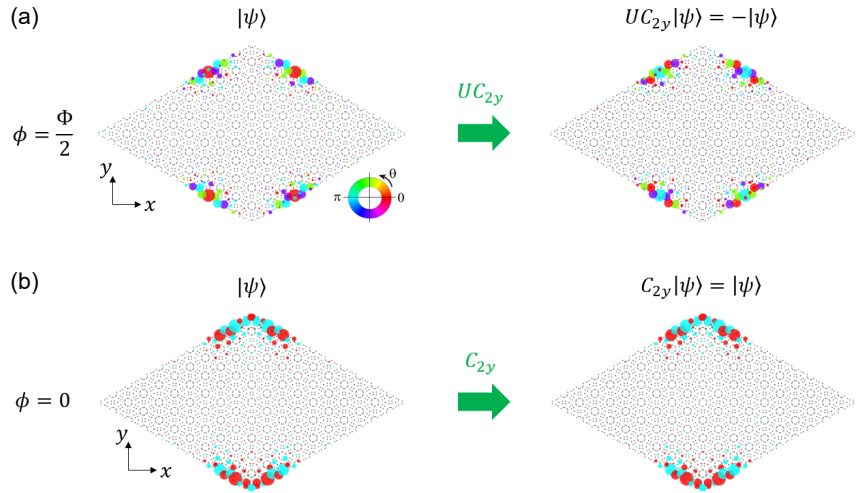

Supplementary Figure 4: (a) Corner state $|\psi\rangle$ at $\phi = \frac{\Phi}{2}$ with $\mathcal{U}C_{2y}$ eigenvalue $-1$, which leads to the nodal structure at the $x = 0$ line. (b) Corner state $|\psi\rangle$ at $\phi = 0$ with $C_{2y}$ eigenvalue $+1$, which leads to the nodal-free structure at the $x = 0$ line.

*3.2. Replica HOTI states at fluxes $\phi = \frac{p}{14}\Phi$ $(p \in \mathbb{Z}; p \neq 7\mathbb{Z})$*

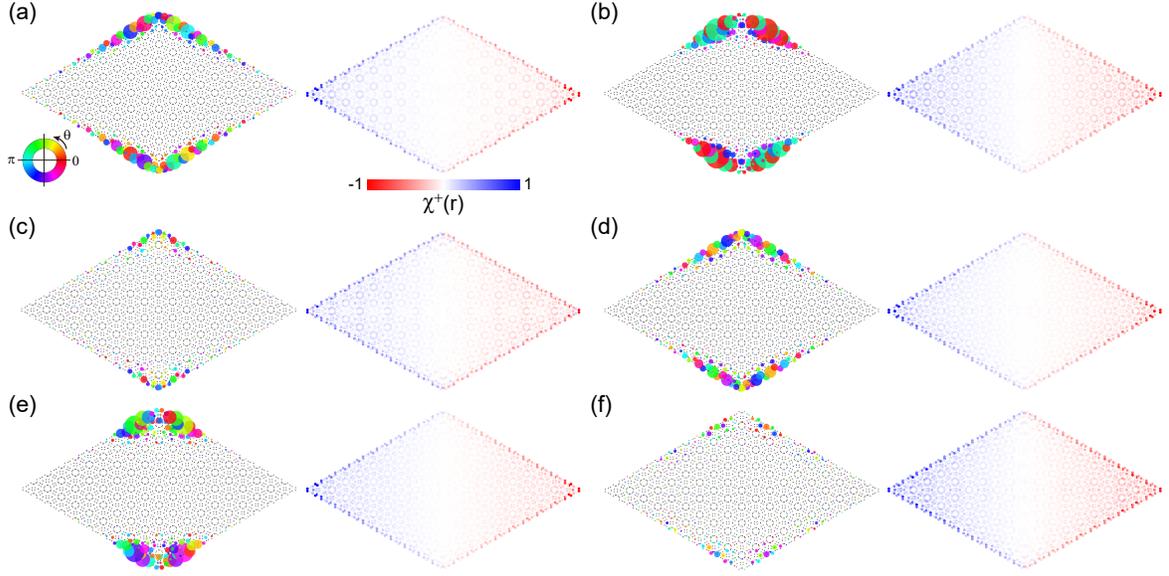

Supplementary Figure 5: (a)-(f) The highest occupied eigenstates and $\mathcal{X}^+(\mathbf{r})$ of replica HOTI states at fluxes $\phi = \frac{p}{14}\Phi$ $(p = 1, 2, \ldots, 6)$. Due to the mirror symmetry about the flux $\phi = \Phi/2$, the eigenstates and markers at $\phi = \frac{p}{14}\Phi$ $(p = 1, 2, \ldots, 6)$ are the same as those at $\phi = \frac{14-p}{14}\Phi$. In (a), the colored circle indicates the phases of eigenstate components. The color scales for $\mathcal{X}^+(\mathbf{r})$ are normalized by the maximum value at each flux.

Supplementary Figure 5 shows the highest occupied states and HOTI markers $\mathcal{X}^+(\mathbf{r})$ at the specific fluxes of the quasi-periodicity $\phi = \frac{p}{14}\Phi$ $(p \in \mathbb{Z}; p \neq 7\mathbb{Z})$. The highest occupied states are found to be the complex-valued corner states and located inside the spectral gap of the bulk (Figs. **2e-j** in the main text), supporting the HOTI states. The HOTI markers $\mathcal{X}^+(\mathbf{r})$ show robust edge-localized features with small permeated values towards the bulk, enabling us to assign them as replica HOTI phases. Here, the composite symmetry $\mathcal{U}_0 C_{2x}$, which is exact for the zero interlayer coupling limit, is responsible for the replica HOTIs.

*3.3. Quantitative analysis of HOTI markers for various states*

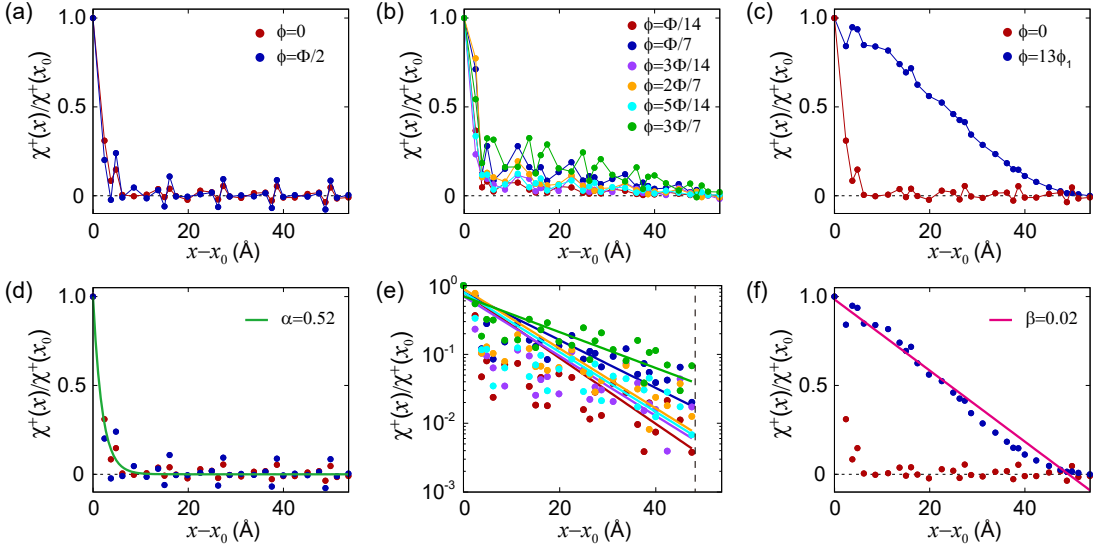

Supplementary Figure 6: (a)-(c) Line profiles of $\mathcal{X}^+(\mathbf{r})$ along the arrow indicated in Supplementary Figure 3(b) for (a) exact HOTIs, (b) replica HOTIs, and (c) trivial state. Here, $x_0$ is the corner position and $\phi_1 = \frac{1}{21000}\Phi$. (d) Line profiles for the exact HOTIs with the fitting function $\exp[-\alpha(x-x_0)]$. (e) Line profiles for the replica HOTIs plotted in log scale. A solid line denotes the fitting function $\propto \exp[-\alpha(x-x_0)]$ with the exponent values $\alpha$ ranging from 0.06 to 0.11, in which the data points are oscillating. The black vertical dashed lines denote the cutoff introduced to avoid plotting zero and negative values in the log scale. (f) Line profile of the trivial state with the fitting function $-\beta(x-x_0) + 0.98$.

We now quantitatively analyze the HOTI marker characteristics for the exact HOTIs, replica HOTIs, and a trivial state (Supplementary Figure 6). The difference between HOTI states and the trivial state manifests in the line profile of the marker $\mathcal{X}^+(\mathbf{r})$: the markers of the HOTI states decay exponentially along the bulk as $\exp[-\alpha(x-x_0)]$ ($\alpha > 0$) [Supplementary Figures 6(d) and 6(e)], whereas the marker of the trivial state at $\phi = 13\phi_1$ is fitted to a linear function $\propto -\beta(x-x_0)$ ($\beta > 0$) [Supplementary Figure 6(f)]. As discussed in the Methods in the main text, the exponential localization from the edge of $\mathcal{X}^+(\mathbf{r})$ for HOTI states fundamentally originates from the action of the projected symmetry operator $P^+C_{2x}P^+$, while the linear delocalization over the entire geometry of $\mathcal{X}^+(\mathbf{r})$ for the trivial state just follows from the form of the marker formula in Eq. (24) in the main text, proportional to the position operator $\hat{X}$. The markers of two exact HOTI states are fitted by the same coefficient $\alpha = 0.52$ [Supplementary Figure 6(d)]. The value of $\alpha$ is reduced to between 0.11 and 0.06 for replica HOTI states [Supplementary Figure 6(e)]. Note that

11

since replica HOTIs can be viewed as disordered versions of exact HOTIs, $\chi^+(\mathbf{r})$ for replica HOTIs exhibit the oscillating behavior.

We additionally present the line profile data for other lines to support that the topological marker of HOTI states is sufficiently localized along the entire edge even in the presence of the symmetry breaking (e.g. at $\phi = \frac{1}{16800}\Phi$; Supplementary Figure 7). The line profile data for various lines in Supplementary Figures 7(c)-(e) show that the topological markers are localized along the entire edge. The log-scale plots indeed show linear behavior along the line and thus the exponential localization of the topological markers [see cyan solid lines in Supplementary Figures 7(f)-(h)]. Note that we introduce the cutoff to avoid plotting zero and negative values of the topological markers.

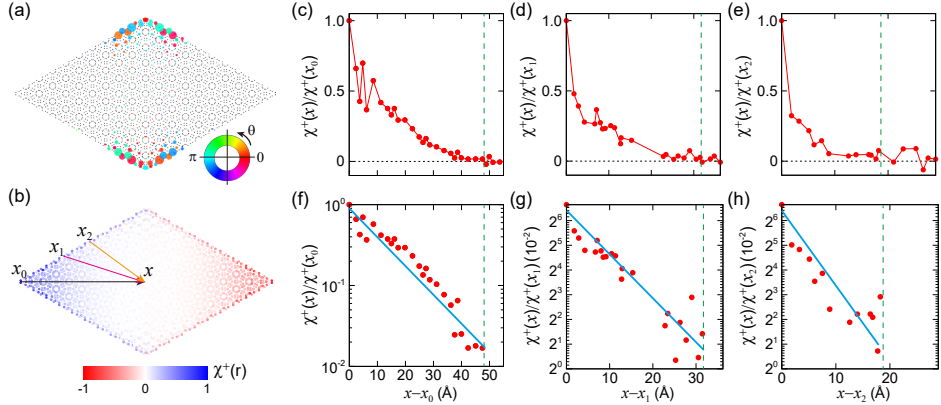

Supplementary Figure 7: (a)-(b) The highest occupied states and $\chi^+(\mathbf{r})$ of the HOTI state for $\phi = \frac{1}{16800}\Phi$. Here, $x_0$, $x_1$ and $x_2$ are positions at the edge. (c)-(e) Line profiles of $\chi^+(\mathbf{r})$ along the lines $\mathbf{x_0 x}$, $\mathbf{x_1 x}$, and $\mathbf{x_2 x}$ indicated in (b). (f)-(h) Same line profiles as (c)-(e) in the log scale. The green dashed lines denote the cutoff, introduced to avoid plotting zero and negative values in the log scale.

## Supplementary Note 4. Local Chern marker: Quantum Hall chiral edge states

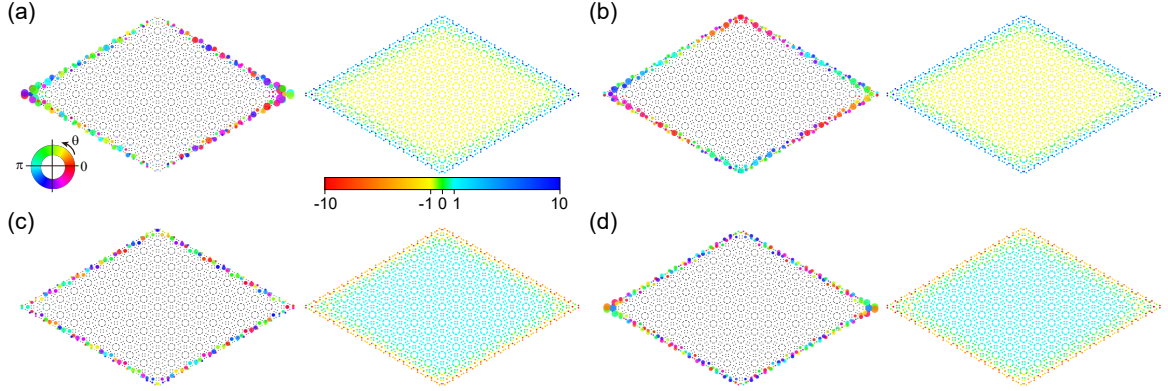

Supplementary Figure 8: (a)-(d) The highest occupied eigenstates and local Chern marker $\mathcal{C}(\mathbf{r})$ of quantum Hall chiral edge states at fluxes $\phi = \frac{q}{42}\Phi$ ($q = 1, 2, 4, 5$). In (a), the colored circle indicates the phases of eigenstate components.

We characterize the states at the discontinuity transitions occuring at fluxes $\phi = \frac{q}{42}\Phi$ ($q \in \mathbb{Z}; q \neq 3\mathbb{Z}$) as quantum Hall chiral edge states by analyzing their eigenstates and local Chern marker $\mathcal{C}(\mathbf{r})$ (Supplementary Figure 8). The eigenstates are localized along the entire edge, which is the characteristic feature of the quantum Hall edge states. The local Chern marker $\mathcal{C}(\mathbf{r})$ indeed reveals nontrivial bulk Chern number where $\mathcal{C}(\mathbf{r})$ is defined by [2]

$$\mathcal{C}(\mathbf{r}) \equiv \langle \mathbf{r} | P[[\hat{X}, P], [\hat{Y}, P]] | \mathbf{r} \rangle = \langle \mathbf{r} | [P\hat{X}P, P\hat{Y}P] | \mathbf{r} \rangle. \tag{17}$$

The local Chern markers $\mathcal{C}(\mathbf{r})$ for $q = 1, 2, 4$, and $5$ of $\phi = \frac{q}{42}\Phi$ show nearly quantized values of $-1, -1, 1$, and $1$ in the bulk region, respectively, which correspond to the bulk Chern numbers $C = -1, -1, 1$, and $1$, respectively.

**Supplementary Note 5. HOTI phases at other large angles**

To address the generalization of the twisted angle, we numerically tested for the next largest two angles. We show the existence of the HOTI states at zero flux in the other large angles. Supplementary Figure 9 shows the corner states at several twist angles $\theta_{m,n} = 21.8°$ $(m=1, n=2), 13.2°$ $(m=2, n=3)$, and $9.4°$ $(m=3, n=4)$. It is noteworthy that the bulk gap is still finite at the twist angles, although the bulk gap is reduced from 9 meV to 0.2 meV [Supplementary Figures 9(a1)-(c1)] as the angle $\theta_{m,n}$ decreases. This guarantees the well-defined corner states inside the spectral bulk gap.

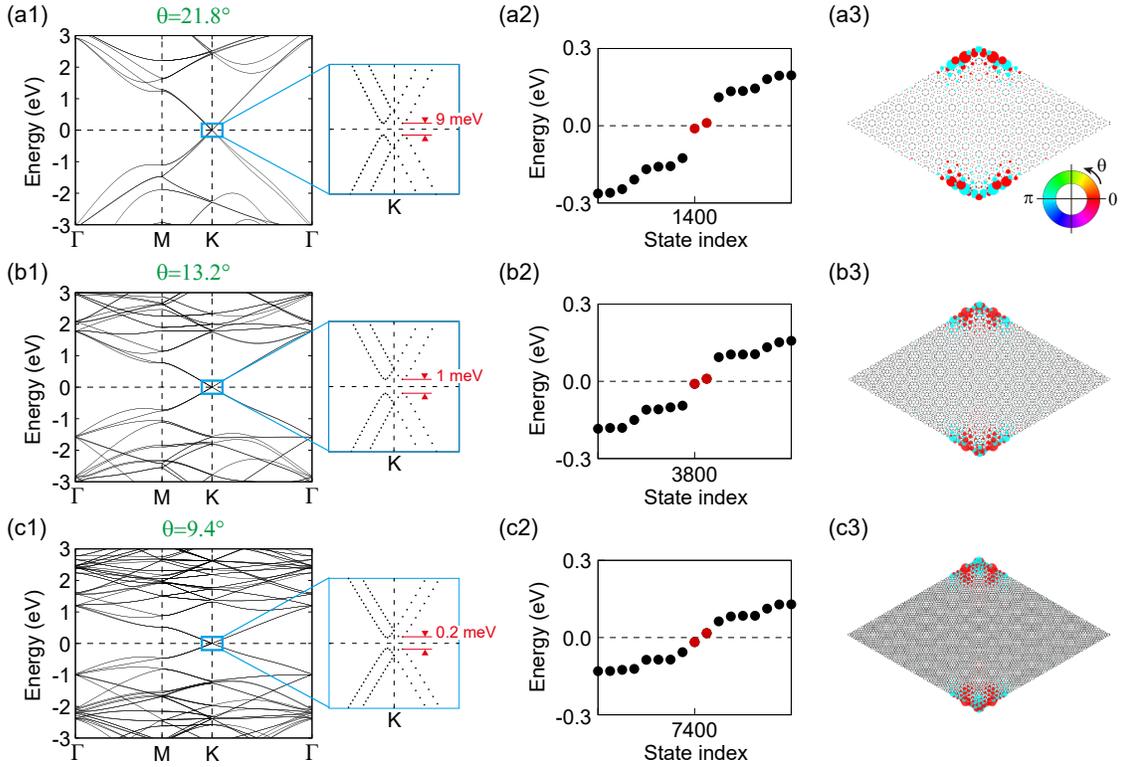

Supplementary Figure 9: **HOTI states at other large angles at zero flux.** (a)-(c) Bulk band structures, energy spectrum in open boundary condition, and real-space distribution of corner states at the twist angle (a) $\theta_{1,2} = 21.8°$, (b) $\theta_{2,3} = 13.2°$, and (c) $\theta_{3,4} = 9.4°$. In (a1)-(c1), the zoomed views of the band structures show the bulk gap $E_g = 9, 1$, and 0.2 meV for $\theta = 21.8°, 13.2°$, and $9.4°$, respectively. For the geometry in the open boundary condition, the $10 \times 10$ unit cell is used for the angles $\theta = 21.8°, 13.2°$, and $9.4°$, which contains 2800, 7600, 14800 atoms, respectively. In (a2)-(c2), red data points indicate corner states residing inside the spectral gap of the bulk.

14

**Supplementary Note 6. Distinct properties of replica HOTIs from exact HOTIs**

We find that the distinct symmetry dependence of the replica HOTIs from the exact HOTIS directly impacts the symmetry bound in the optical and transport properties. As we elaborated in the manuscript, the exact HOTIs respect symmetries, while the replicas do not. This loose dependence of the replica HOTIs on the protecting symmetries directly leads to the following optical and transport properties.

Supplementary Figure 10 shows the non-zero values of the degree of circular polarization $\eta$ in replica HOTIs, in stark contrast to the zero value of $\eta$ in exact HOTIs (details of the calculations can be found in the caption). We attribute this difference to the non-zero finite values of the position operator matrix elements between two corner states $\langle\psi_2|\hat{X}|\psi_1\rangle$ in replica HOTIs, which are allowed in the absence of protecting symmetry. Furthermore, we expect that the difference in the matrix elements can lead to distinctions in other associated observable quantities between the exact and replica HOTI phases.

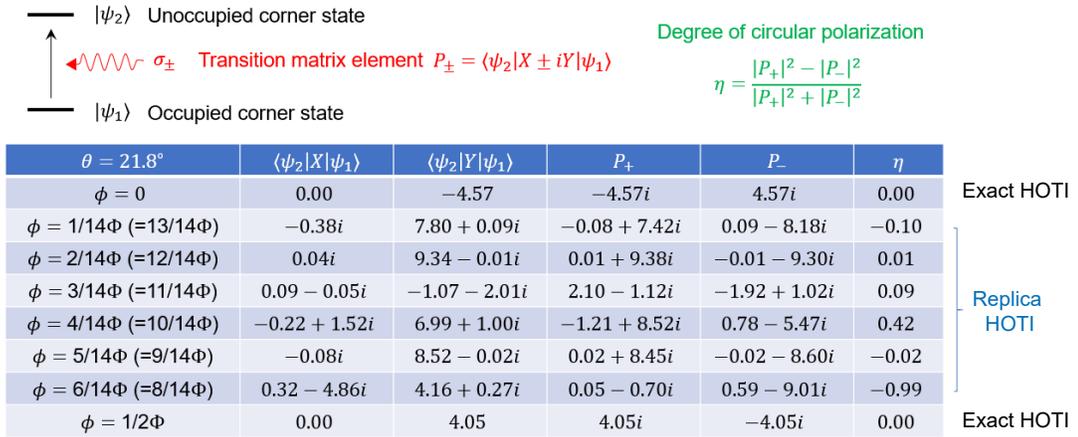

| $\theta = 21.8°$ | $\langle\psi_2|X|\psi_1\rangle$ | $\langle\psi_2|Y|\psi_1\rangle$ | $P_+$ | $P_-$ | $\eta$ | |
|---|---|---|---|---|---|---|
| $\phi = 0$ | 0.00 | $-4.57$ | $-4.57i$ | $4.57i$ | 0.00 | Exact HOTI |
| $\phi = 1/14\Phi$ (=13/14$\Phi$) | $-0.38i$ | $7.80 + 0.09i$ | $-0.08 + 7.42i$ | $0.09 - 8.18i$ | $-0.10$ | |
| $\phi = 2/14\Phi$ (=12/14$\Phi$) | $0.04i$ | $9.34 - 0.01i$ | $0.01 + 9.38i$ | $-0.01 - 9.30i$ | $0.01$ | |
| $\phi = 3/14\Phi$ (=11/14$\Phi$) | $0.09 - 0.05i$ | $-1.07 - 2.01i$ | $2.10 - 1.12i$ | $-1.92 + 1.02i$ | $0.09$ | Replica HOTI |
| $\phi = 4/14\Phi$ (=10/14$\Phi$) | $-0.22 + 1.52i$ | $6.99 + 1.00i$ | $-1.21 + 8.52i$ | $0.78 - 5.47i$ | $0.42$ | |
| $\phi = 5/14\Phi$ (=9/14$\Phi$) | $-0.08i$ | $8.52 - 0.02i$ | $0.02 + 8.45i$ | $-0.02 - 8.60i$ | $-0.02$ | |
| $\phi = 6/14\Phi$ (=8/14$\Phi$) | $0.32 - 4.86i$ | $4.16 + 0.27i$ | $0.05 - 0.70i$ | $0.59 - 9.01i$ | $-0.99$ | |
| $\phi = 1/2\Phi$ | 0.00 | 4.05 | $4.05i$ | $-4.05i$ | 0.00 | Exact HOTI |

Supplementary Figure 10: **Various matrix elements between the two corner states $|\psi_1\rangle$ and $|\psi_2\rangle$ of the exact ($\phi = 0, \frac{1}{2}\Phi$) and replica HOTIs ($\phi = \frac{p}{14}\Phi, p \in \mathbb{Z}; p \neq 7\mathbb{Z}$).** Here, we only consider the matrix elements between the occupied corner state $|\psi_1\rangle$ and the unoccupied corner state $|\psi_2\rangle$ for simplicity. We consider transition matrix element for the left ($\sigma_+$) and right ($\sigma_-$) circularly polarized light absorption as $P_\pm = \langle\psi_2|X \pm iY|\psi_1\rangle$. The degree of circular polarization is defined by the difference between left and right circularly polarized light absorption as $\eta = \frac{|P_+|^2 - |P_-|^2}{|P_+|^2 + |P_-|^2}$.

Furthermore, regarding the transport properties, the replica HOTI phases show distinct tunneling behaviors of corner states from exact HOTIs as shown in Supplementary Figure 11. Exact HOTIs show the oscillatory behavior of the energy splitting between the corner states.

15

The energy splitting is given by $\Delta E^2 = 4K\sqrt{\frac{S_0}{2\pi}}e^{-S_0}|\cos(\gamma)|$ [3] where $K$ is the constant determinant, $S_0$ is the action of the instanton, and $\gamma$ is the geometric phase difference arising from the intrinsic Berry phase of the HOTI phase. In contrast, replica HOTIs exhibit nearly degenerate energies of the two corner states as a function of the gate voltage. We attribute such nearly flat dispersion to the decoupling of the two corner states in the absence of the protecting symmetry. We leave the detailed mechanism of this phenomenon as a future work.

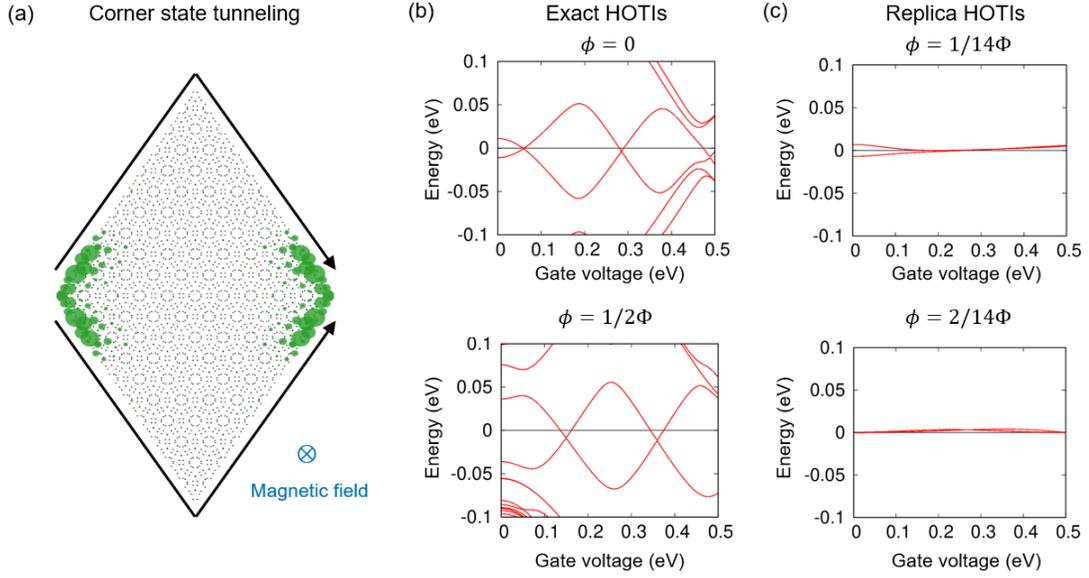

Supplementary Figure 11: (a) The wave function distribution (green dots) of the corner states in TBG. Black arrows illustrate the tunneling paths that connect the corner states. The two paths together form a complete loop along the edge of the HOTI. (b)-(c) The evolution of corner states energies as a function of the gate voltage of (b) the exact HOTIs ($\phi = 0$ and $\phi = \frac{1}{2}\Phi$) and (c) the replica HOTIs ($\phi = \frac{1}{14}\Phi$ and $\phi = \frac{2}{14}\Phi$). Here, the zero energy is defined as the average value of the two corner states at zero gate voltage.


\fi

\end{document}